\documentclass[twocolumn,times]{aastex63}

\usepackage{graphicx}
\usepackage[FIGTOPCAP]{subfigure}
\usepackage{amsmath,booktabs,threeparttable,url,bm}
\usepackage{CJK}
\usepackage{natbib}

\newcommand{\tachar}[1]{
\setbox4=\hbox{\ }
\setbox3=\hbox{#1}
\hbox{#1}
\kern -\wd3 \kern -\wd4
\raise 0.45\ht3 \hbox{ \vrule width \wd3 height 0.5pt}
}
\newcommand{\tachard}[1]{
\setbox4=\hbox{\ }
\setbox3=\hbox{#1}
\hbox{#1}
\kern -\wd3 \kern -\wd4
\raise 0.35\ht3 \hbox{ \vrule width \wd3 height 0.5pt}
}
\newcommand{\rquad}{\tachar{I}}

\newcommand{\rquaddot}{\tachard{\"{I}}}

\newcommand{\rads}[1]{rad~s$^{-1}$}
\newcommand{\gcm}[1]{g~cm$^{-3}$}
\newcommand{\dyncm}[1]{dyn~cm$^{-2}$}
\newcommand{\kms}[1]{km~s$^{-1}$}
\newcommand{\tw}[1]{$T/|W|$}
\newcommand{\bvfreq}[1]{Brunt-V\"{a}is\"{a}l\"{a}}

\newcommand{\flash}[1]{\textsc{FLASH}}
\newcommand{\sphynx}[1]{\textsc{SPHYNX}}

\graphicspath{{figures/}}

\received{\today}

\shorttitle{GW features in rotating CCSNe}
\shortauthors{Hsieh et al.}

\begin{document}
\begin{CJK*}{UTF8}{bsmi}

\title{A New Kilohertz Gravitational-Wave Feature from Rapidly Rotating Core-Collapse Supernovae}

\newcommand*{\NTHUP}{Department of Physics, National Tsing Hua University, Hsinchu 30013, Taiwan}
\newcommand*{\NTHUA}{Institute of Astronomy, National Tsing Hua University, Hsinchu 30013, Taiwan}
\newcommand*{\CICA}{Center for Informatics and Computation in Astronomy, National Tsing Hua University, Hsinchu 30013, Taiwan}
\newcommand*{\BASEL}{Center for Scientific Computing - sciCORE, Universit\"{a}t Basel, Klingelbergstrasse 82, CH-4056 Basel, Switzerland}
\newcommand*{\CTC}{Center for Theory and Computation, National Tsing Hua University, Hsinchu 30013, Taiwan}
\newcommand*{\NCTS}{Physics Division, National Center for Theoretical Sciences, National Taiwan University, Taipei 10617, Taiwan}
\newcommand*{\NTU}{Department of Physics, National Taiwan University, Taipei 10617, Taiwan}

\author[0000-0002-8947-723X]{He-Feng Hsieh}
\affiliation{\NCTS} \affiliation{\NTU} \affiliation{\NTHUA} \affiliation{\CICA}

\author[0000-0003-3546-3964]{Rub\'{e}n Cabez\'{o}n}
\affiliation{\BASEL}

\author[0009-0000-0674-7592]{Li-Ting Ma (馬麗婷)}
\affiliation{\NTHUA} \affiliation{\CICA} \affiliation{\NTHUP} 

\author[0000-0002-1473-9880]{Kuo-Chuan Pan (潘國全)}
\affiliation{\NCTS} \affiliation{\NTHUA} \affiliation{\CICA} \affiliation{\NTHUP}  \affiliation{\CTC}


\begin{abstract}

We present self-consistent three-dimensional core-collapse supernova simulations of a rotating $20M_\odot$ progenitor model with various initial angular velocities from $0.0$ to $4.0$~\rads{} using a smoothed particle hydrodynamics code, \sphynx{}, and a grid-based hydrodynamics code, \flash{}. 
We identify two strong gravitational-wave features, with peak frequencies of $\sim300$~Hz and $\sim1.3$~kHz in the first $100$~ms postbounce. 
We demonstrate that these two features are associated with the $m=1$ deformation from the proto-neutron star (PNS) modulation induced by the low-\tw{} instability, regardless of the simulation code.
The $300$~Hz feature is present in models with an initial angular velocity between $1.0$ and $4.0$~\rads{}, while the $1.3$~kHz feature is present only in a narrower range, from $1.5$ to $3.5$~\rads{}. 
We show that the $1.3$~kHz signal originates from the high-density inner core of the PNS, and the $m=1$ deformation triggers a strong asymmetric distribution of electron anti-neutrinos.
In addition to the $300$~Hz and $1.3$~kHz features, we also observe one weaker but noticeable gravitational-wave feature from higher-order modes in the range between $1.5$ and $3.5$~\rads{}. Its peak frequency is around $800$~Hz initially and gradually increases to $900-1000$~Hz.
Therefore, in addition to the gravitational bounce signal, the detection of the $300$~Hz, $1.3$~kHz, the higher-order mode, and even the related asymmetric emission of neutrinos, could provide additional diagnostics to estimate the initial angular velocity of a collapsing core. 

\end{abstract}


\section{INTRODUCTION}

Detection of gravitational waves (GWs) together with neutrino emissions from nearby core-collapse supernovae (CCSNe) will place meaningful constraints on the supernova engine(s) and the nuclear equation of state. In the past decade, tremendous efforts have been dedicated to producing GW waveform predictions from CCSNe \citep{2021Natur.589...29B}, including non-rotating progenitors \citep{2018ApJ...865...81O, 2020ApJ...901..108V, 2020PhRvD.102b3027M, 2023PhRvD.107d3008M}, light massive stars \citep{2015ApJ...801L..24M, 2019MNRAS.485.3153B}, rotating progenitors \citep{2018MNRAS.475L..91T, 2019MNRAS.486.2238A, 2020MNRAS.494.4665P, 2021MNRAS.502.3066S, 2021MNRAS.508..966T}, failed supernovae \citep{2021ApJ...914..140P, 2021ApJ...911....6I}, magnetized CCSNe \citep{2020ApJ...896..102K, 2021MNRAS.503.4942O, 2022MNRAS.510.5535J, 2023MNRAS.522.6070P}, and pulsational pair-instability supernovae \citep{2021MNRAS.503.2108P, 2022MNRAS.512.4503R}. However, these studies suggest that only galactic CCSNe or fast-rotating progenitors have a reasonable chance of being detected by the ongoing O4 (or the next O5) observational run of the Advanced LIGO, Advanced Virgo, and KAGRA network \citep{2020LRR....23....3A, 2021PhRvD.104j2002S}. 

To prepare ourselves for future GW detection from a nearby CCSN, we have to understand the GW features from CCSNe through (magneto-)hydrodynamical simulations as those described above, and/or GW asteroseismology \citep{2018MNRAS.474.5272T, 2018ApJ...861...10M, 2021PhRvD.104l3009S}. 
\cite{2017PhRvD..95f3019R} conducted $1824$ axisymmetric general-relativistic (GR) hydrodynamical simulations that include a wide range of angular velocities of a $12 M_\odot$ progenitor. However, due to limits on computational resources, \cite{2017PhRvD..95f3019R} calculated only up to the first $\sim50$~ms after the core bounce, focused on the bounce signals, and conducted the calculations exclusively with the s12 progenitor. 
\cite{2019ApJ...878...13P, 2021ApJ...914...80P} conducted a similar analysis for different progenitor masses and performed longer simulations during the accretion phase. However, in their latest work, the analysis is done with approximately fifty 2D simulations and only four 3D simulations. 
Three-dimensional fluid instabilities, such as standing acceleration shock instability (SASI; \citealt{2003ApJ...584..971B}) could play an important role in the GW features \citep{2016ApJ...829L..14K}, and the spiral modes of SASI can only be developed in 3D simulations. Therefore, extending the previous works to full 3D calculations is crucial and necessary to investigate the realistic GW features from CCSNe.

Unlike simulations of binary neutron star mergers, it is important to highlight that gravitational waveforms produced from CCSN simulations rely on grid-based hydrodynamics codes. Additionally, the majority of these grid-based hydrodynamics codes adopt either Cartesian coordinates \citep{2018ApJ...865...81O, 2020ApJ...896..102K, 2021ApJ...914..140P, 2021MNRAS.502.3066S} or spherical coordinates \citep{2019MNRAS.485.3153B, 2021MNRAS.503.2108P, 2021MNRAS.508..966T}.
Spherical grids require an inner boundary condition and/or a fixed proto-neutron star (PNS) center. Such limitations might restrict the GW emissions from regions close to the coordinate center or the inner boundary.  
On the other hand, Cartesian grids have no inner boundary conditions at the coordinate center but can generate $m = 4$ perturbations \citep[e.g.,][]{ 2007PhRvL..98z1101O} and therefore, artificial GW emissions from these grid effects may pollute the GW features. 
Furthermore, sound waves might oscillate and bounce back between grid-refinement boundaries, which might also induce artificial GW emissions.
An alternative to avoid these artifacts is to use a pure meshless Lagrangian method, such as smoothed particle hydrodynamics (SPH), which does not suffer from these grid effects. Furthermore, GW emissions can be recovered directly from the particle distribution and physical magnitudes carried by each particle, without time derivatives involved, producing clean and reliable waveforms \citep{1993ApJ...416..719C}.

An additional reason for using an SPH code in this work is its intrinsic conservation properties and, in particular, angular momentum conservation. Given that we are simulating from slow to fast rotators, angular momentum conservation and transfer are key aspects to be confident in our results. 
SPH codes are constructed so that their momentum and energy equations are pairwise equivalent between particles and their neighbors \citep{2005RPPh...68.1703M}. This, theoretically, ensures conservation to machine precision. Note, nevertheless, that in real simulations, this conservation is somewhat degraded by the inclusion of self-gravity, but it is still maintained at consistently high accuracy.

Therefore, in this work, we investigate the GW features of a wide range of initial angular velocities of a $20 M_\odot$ progenitor, using the SPH code \sphynx{} \citep{2017A&A...606A..78C, 2022A&A...659A.175G}. A subset of these simulations was also repeated with the grid-based adaptive mesh refinement (AMR) code, \flash{} \citep{2000ApJS..131..273F, 2008PhST..132a4046D}, to ensure that our conclusions are code-independent, and when differences appear, we could understand them within the framework of the usage of different hydrodynamics solvers.

In the following section, we introduce the numerical methods that we employ. Namely, the hydrodynamics codes, the physics included in them, the initial setup including a list of all calculated models, and the extraction methods of the GW emissions and the spherical harmonic modes. Section~\ref{sec:results} presents our results, focusing on the features in the GW emissions, their dependence on the initial angular velocity, and their potential observability. Finally, Section~\ref{sec:conclusion} offers a summary and conclusion of the results.

\section{NUMERICAL METHODS and MODELS} \label{sec:numer_method}

We describe the numerical codes and the corresponding setup of our simulations in Section~\ref{sec:numer_code}. In Section~\ref{sec:init_setup}, we present the initial conditions of the investigated supernova progenitor and the rotation setup. Finally, we describe the analysis methods for obtaining the GW emissions and spherical harmonic coefficients in Sections~\ref{sec:method_gw} and \ref{sec:method_sphharm}, respectively.

\subsection{Hydrodynamics Codes} \label{sec:numer_code}

\sphynx{}\footnote{\url{https://astro.physik.unibas.ch/sphynx}} is a state-of-the-art SPH code, with accurate gradient evaluation \citep{2012A&A...538A...9G, 2015MNRAS.448.3628R,2022A&A...659A.175G}, pairing-resistant interpolating kernels \citep{2008JCoPh.227.8523C}, generalized volume elements \citep{2013MNRAS.428.2840H, 2013ApJ...768...44S}, adaptive artificial viscosity via switches \citep{2010MNRAS.405.1513R}, and adaptive spatial and temporal resolution. To our knowledge, this is currently the only SPH code that can simulate CCSNe with spectral neutrino treatment.

In the context of CCSN, \sphynx{} was used for the development and comparison of a spectral neutrino leakage scheme \citep{2014A&A...568A..11P}. It was also used in a code-comparison work \citep{2018A&A...619A.118C}, where it showed particularly good agreement with other Eulerian codes, such as \flash{}, at least for the collapse phase and the first $50$~ms postbounce, and for a series of CCSN simulations with several physics implementations and different progenitors. 

The implementation of neutrino treatment in \sphynx{} is based on the Isotropic Diffusion Source Approximation \citep[IDSA;][]{2009ApJ...698.1174L} and is coupled with a parametrized deleptonization \citep{2005ApJ...633.1042L} for the collapse phase. Self-gravity is calculated using a Barnes-Hut algorithm on an octree, and it also includes an effective GR potential correction \citep[Case A in][]{2006A&A...445..273M}, adopted to replace the monopole term of the Newtonian gravitational potential. Finally, we use the LS220 equation of state from \url{stellarcollapse.org} \citep{1991NuPhA.535..331L, 2010CQGra..27k4103O} in all of our simulations. For further details on the coupling of all these physics modules, we refer the reader to Section~2.4 in \cite{2018A&A...619A.118C}.

For comparisons, we additionally perform two simulations with the grid-based Eulerian hydrodynamics code \flash{}\footnote{\url{https://flash.rochester.edu}} 
that includes the IDSA for neutrino transport \citep{2016ApJ...817...72P, 2018ApJ...857...13P, 2019JPhG...46a4001P, 2021ApJ...914..140P}. 
We also use the same effective GR potential and LS220 equation of state.

\subsection{Initial Setup} \label{sec:init_setup}

In this work, we employ the same $20 M_\odot$ spherically symmetric progenitor star with solar metallicity as the initial conditions for all simulations. The initial radial profiles of density, temperature, electron fraction, and radial velocity are adopted from the 1D, s20 model provided by \cite{2007PhR...442..269W}. 
The initial 3D particle distribution for the SPH calculations was achieved following the original 1D density profile but with a random angular arrangement. Next, the resulting 3D distribution was relaxed, allowing the particles to move only tangentially (i.e., at a fixed radius), which smoothed out spurious clumps of particles and provided clean, smooth initial profiles. 
A small external pressure ($P_\mathrm{ext} = 1.8 \times 10^{22}$~\dyncm{}) was added to all particles, similar to the pressure exerted by the outer layers of the star, which are not included in the simulation. This prevented particles in the very low-density region from experiencing an artificially high gradient of pressure and escaping from the system. Note that $P_\mathrm{ext}$ is only relevant to a very thin layer of particles in the outer region of the simulated section of the star, as the pressure profile ranges between $10^{24}-10^{33}$~\dyncm{} for the vast majority of the domain.
Regarding neutrino transport, we use $20$ energy bins, logarithmically spaced from $3$ until $300$~MeV for the electron-flavor neutrinos. The heavier $\mu$ and $\tau$ neutrinos and anti-neutrinos are considered as a single species that cools the system via a leakage scheme.

Finally, an artificial tangential velocity was imposed to induce rotation. We assigned the initial rotational profile as follows,
\begin{equation}
    \Omega (r) = \dfrac{\Omega_{0}}{1+\left(\frac{r}{A}\right)^2}, \label{eq:rot} 
\end{equation}
where $\Omega$ is the angular velocity as a function of the spherical radius $r$, $A = 1.03376 \times 10^{8}$~cm is a characteristic length that is taken from the fitting formula provided in \cite{2019ApJ...878...13P} (see Figure~2 therein), and $\Omega_{0}$ is the angular velocity at the center of the star, of which we investigate several values ranging from $0.0$ to $4.0$~\rads{}. 

We perform the SPH simulations with three different resolutions, which contain $200$k, $675$k, and $1600$k particles that cover the central $10^4$~km of the progenitor star. The corresponding resolutions within $10$~km in the radius are about $600$, $400$, and $300$~m after the core bounce. The primary analysis and results presented in this paper are based on the high-resolution simulations with $1600$k particles. Furthermore, we used the simulations with $675$k particles to investigate a larger parameter space of $\Omega_{0}$.

\begin{deluxetable*}{cccccccc}
\tablewidth{0pt} 
\label{table:models}
\tablecaption{Summary of the models. 
Columns from left to right show the model name, numerical code, initial angular velocity ($\Omega_0$), number of particles, peak frequency of the kHz signal, the value of \tw{} at the initial time, and the highest spatial resolution at $20$~ms postbounce ($\Delta x_\mathrm{min, 20ms}$).}
\tablehead{
Model &
Code &
$\Omega_0$ &
\# of particles &
$f_\mathrm{peak}$ &
\tw{}$_\mathrm{init}$ &
$\Delta x_\mathrm{min, 20ms}$
\\ 
& & [\rads{}] & [$10^3$] & [kHz] & $[10^{-3}]$ & [m]
}
\startdata
S00   &  SPHYNX  &  $0.0$  &   $675$  &      -   &  $4.84 \times 10^{-10}$  &  $372.3$ \\
S05   &  SPHYNX  &  $0.5$  &   $675$  &      -   &  $0.12$  &  $372.4$ \\
S10   &  SPHYNX  &  $1.0$  &   $675$  &      -   &  $0.48$  &  $374.4$ \\
S15   &  SPHYNX  &  $1.5$  &   $675$  &      -   &  $1.09$  &  $378.5$ \\
S20   &  SPHYNX  &  $2.0$  &   $675$  &  $1.35$  &  $1.94$  &  $383.7$ \\
S25   &  SPHYNX  &  $2.5$  &   $675$  &  $1.29$  &  $3.02$  &  $389.3$ \\
S30   &  SPHYNX  &  $3.0$  &   $675$  &  $1.29$  &  $4.35$  &  $395.4$ \\
S35   &  SPHYNX  &  $3.5$  &   $675$  &  $1.34$  &  $5.93$  &  $403.0$ \\
S40   &  SPHYNX  &  $4.0$  &   $675$  &      -   &  $7.74$  &  $410.9$ \\
S30L  &  SPHYNX  &  $3.0$  &   $200$  &  $1.28$  &  $4.35$  &  $591.9$ \\
S20H  &  SPHYNX  &  $2.0$  &  $1600$  &  $1.38$  &  $1.94$  &  $288.1$ \\
S30H  &  SPHYNX  &  $3.0$  &  $1600$  &  $1.12-1.32$  &  $4.36$  &  $296.9$ \\
\hline
\hline
F20   &  FLASH   &  $2.0$  &     -    &  $1.01-1.48$  &  $1.89$  &  $488.3$ \\
F30   &  FLASH   &  $3.0$  &     -    &  $1.21-1.39$  &  $4.25$  &  $488.3$
\enddata
\end{deluxetable*}

The Cartesian grid setup in our \flash{} simulations closely follows the setup described in \cite{2021ApJ...914..140P}. Therefore, we just give here a brief review of our setup. We use a three-dimensional simulation box that covers the inner $10^4$~km of the CCSN progenitor, and employ nine levels of AMR, which yield a maximum spatial resolution of $488$~m at the highest AMR level.
The central $r < 120$~km sphere has the highest spatial resolution, while we reduce the AMR level as we move farther away from the center of the progenitor to save computing time. This results in an effective angular resolution $\sim 0^\circ.2 - 0^\circ.4$. For the outer boundary conditions, we use a power-law profile that depends on the spherical radius. 
In addition, we adopt the same $20$ neutrino energy bins as in \sphynx{}, spaced logarithmically from $3$ to $300$~MeV for the electron flavor neutrinos and a leakage scheme for the $\mu$ and $\tau$ neutrinos. We use the same s20 progenitor from \cite{2007PhR...442..269W} and take the same rotational profile as in Equation~(\ref{eq:rot}) with $\Omega_0 = 2$ and $3$~\rads{}.

Table \ref{table:models} summarizes the relevant information for all our models. Letters F and S in the model name denote the code used to perform the simulation, \flash{} or \sphynx{}, respectively. 
The number in the model names shows the adopted value of the initial central angular velocity, $\Omega_0$.
Finally, L and H stand for low and high resolution, respectively, for \sphynx{} models. The column $f_\mathrm{peak}$ shows the resulting peak GW frequency around kHz in the time interval $t_\mathrm{pb} = 10-100$~ms. We provide a range of peak frequencies for models that exhibit noticeable variations in their peak frequencies over time.
The sixth column of Table~\ref{table:models} indicates the ratio between rotational energy and gravitational energy, \tw{}, calculated for the regions with $\rho \ge 10^6$~\gcm{} at the initial time.
The last column $\Delta x_\mathrm{min, 20ms}$ shows the highest resolution for each model, which is defined as the smallest smoothing length for \sphynx{} models or the smallest cell size for \flash{} models, at $20$~ms postbounce.

\subsection{Extracting the Gravitational Waves} \label{sec:method_gw}

In order to extract the GW emissions in \sphynx{}, we adopt the transverse-traceless gauge to cover the far zone of the source. Hence, we have only two polarizations of the amplitude of the GWs:
\begin{align}
h_+&=\frac{1}{D}\frac{G}{c^4}\left(\rquaddot_{\theta \theta}-\rquaddot_{\phi \phi}\right)\label{hplus},\\
h_\times&=\frac{1}{D}\frac{G}{c^4}\rquaddot_{\theta \phi}\label{hcross}\,,
\end{align}
where $G$ is gravitational constant, $c$ is the speed of light, and $D$ is the distance to the source. The reduced quadrupole ($\rquad_{lm}$) is defined in Cartesian coordinates as,
\begin{equation}
\rquad_{lm}=\int \rho \left( x_lx_m-\frac{1}{3}\delta_{lm}x_kx^k\right)d^3x\,.
\label{rquad}
\end{equation}
where $l$ and $m$ are components of the position vector $\mathbf{x}$, $\delta_{lm}$ is a $\delta$-Kronecker, and $\rho$ is the local density.

The components of the reduced quadrupole in spherical coordinates are related to the Cartesian by \citep{1997PThPS.128..183O},
\begin{align}
\rquaddot_{\theta\theta}&=\left(\rquaddot_{xx}\cos^2\phi+\rquaddot_{yy}\sin^2\phi+\rquaddot_{xy}\sin 2\phi\right)\cos^2\theta\nonumber\\
&\qquad+\rquaddot_{zz}\sin^2\theta-\left(\rquaddot_{xz}\cos\phi+\rquaddot_{yz}\sin\phi\right)\sin 2\theta\label{quadsf1}\,,\\
\rquaddot_{\phi\phi}&=\rquaddot_{xx}\sin^2\phi+\rquaddot_{yy}\cos^2\phi-\rquaddot_{xy}\sin 2\phi\label{quadsf2}\,,\\
\rquaddot_{\theta\phi}&=-\frac{1}{2}\left(\rquaddot_{xx}-\rquaddot_{yy}\right)\cos\theta\sin 2\phi+\rquaddot_{xy}\cos\theta\cos 2\phi\nonumber\\
&\qquad-\left(\rquaddot_{xz}\sin\phi-\rquaddot_{yz}\cos\phi\right)\sin\theta\label{quadsf3}\,.
\end{align}
Therefore, to obtain the GW waveforms we compute the six non-zero components of $\rquaddot_{lm}$ in Cartesian coordinates and then transform them into spherical coordinates using Equations~(\ref{quadsf1}), (\ref{quadsf2}), and (\ref{quadsf3}). Finally, we substitute these components into Equations~(\ref{hplus}) and (\ref{hcross}) to obtain the final waveforms.

There are different methods for evaluating the components of $\rquaddot_{lm}$. Nevertheless, time derivatives cause numerical difficulties due to two main reasons: the numerical noise introduced by the discretization and the magnification of the high-frequency components of the noise. To avoid these problems, we opted to use the method proposed by \cite{1993ApJ...416..719C}, which takes advantage of the Lagrangian nature of SPH and is similar to the stress formula of \cite{1990ApJ...351..588F}. 

A discretized version of Equation~(\ref{rquad}) is
\begin{equation}
\rquad_{lm}=\sum_im^i\left[x_l^ix_m^i-\frac{1}{3}\delta_{lm}\mathbf{r}^i\cdot\mathbf{r}^i\right]\,,
\label{rquaddisc}
\end{equation}
where the subscripts refer to the three Cartesian components, the superscripts label the SPH particles, $m^i$ is the mass of each particle, and the summation is over all the particles of the simulation. Taking the second time derivative of Equation~(\ref{rquaddisc}) is straightforward, obtaining
\begin{align}
\rquaddot_{lm}&=\frac{2}{3}\sum_im^i\left[2v_l^iv_m^i+a_l^ix_m^i+x_l^ia_m^i\right.\nonumber\\
&\qquad+\left.\delta_{lm}\left(v_l^iv_m^i+x_l^ia_m^i-\mathbf{v}^i\cdot\mathbf{v}^i-\mathbf{r}^i\cdot\mathbf{a}^i\right)\right]\,. \label{rquaddotdisc}
\end{align}
The $x$, $v$, and $a$ terms are the components of the position, velocity, and acceleration vectors of particle $i$, respectively. Using Equation~(\ref{rquaddotdisc}) in Equations~(\ref{hplus})$-$(\ref{quadsf3}) we can find the amplitude of both polarizations of the GW emissions directly from magnitudes calculated by the SPH code without having to compute explicit second-time derivatives.

\flash{} also uses the quadrupole formula to extract the GW emissions (Equations~\ref{hplus}$-$\ref{quadsf3}), but unlike \sphynx{}, the GW calculations in \flash{} follow the strain formulation to compute the first time derivative of the quadrupole moment. The second time derivative of the quadrupole moment is evaluated by a finite difference method via post-processing \citep{2018ApJ...857...13P, 2021ApJ...914..140P}.

\subsection{Spherical Harmonic Mode Analysis} \label{sec:method_sphharm}

It is known that, for a moderately rotating core, the low-\tw{} instability could develop after the core bounce, which induces an $m = 1$ or $m = 2$ deformation that results in quasi-sinusoidal oscillations of the GW emissions \citep{2005ApJ...625L.119O, 2020MNRAS.493L.138S}. To investigate the relationship between the GW signals and the deformation induced by the low-\tw{} instability, we apply a spherical harmonic decomposition to the resulting fluid distributions. 
Following \citet{2012ApJ...759....5B}, we evaluate the coefficient of each mode using
\begin{equation}
a_{lm} = \frac{ (-1)^{|m|} }{ \sqrt{4\pi(2l + 1)} } 
         \frac{\sum w R(\theta,\phi) Y_l^m(\theta,\phi)}
              {\sum w},
\label{eq:sphharm}
\end{equation}
where $R$ is the distance from the PNS center, and $w$ is a weighting function taken as the volume. For \sphynx{}, we set the weighting function to the associated volume of the SPH particles, $w = m^i / \rho^i$, where $m^i$ and $\rho^i$ are the mass and density carried by the SPH particles, respectively. 
The orthonormal harmonic basis functions, $Y_l^m$, are expressed as
\begin{equation}
Y_l^m(\theta, \phi) = 
  \begin{cases} 
    \sqrt{2} N_l^m P_l^m(\cos\theta) \cos m\phi&            m>0, \\
    N_l^0 P_l^0(\cos\theta) &                               m=0, \\
    \sqrt{2} N_l^{|m|} P_l^{|m|}(\cos\theta) \sin |m|\phi&  m<0,
  \end{cases}
\end{equation}
where
\begin{equation}
N_l^m = \sqrt{\frac{2l+1}{4\pi}\frac{(l-m)!}{(l+m)!}}
\end{equation}
and $P_l^m(\cos\theta)$ is the associated Legendre polynomial. 

In this work, we focus on the dipole mode ($l = 1$, \mbox{$m = 1$)} and quadrupole mode ($l = 2, m = 2$), which represent the $m = 1$ and $m = 2$ deformations, respectively.

\section{RESULTS} \label{sec:results}
\subsection{Dynamics Overview} \label{sec:dynamic_overview}

\begin{figure*}
    \includegraphics[width=1.0\textwidth]{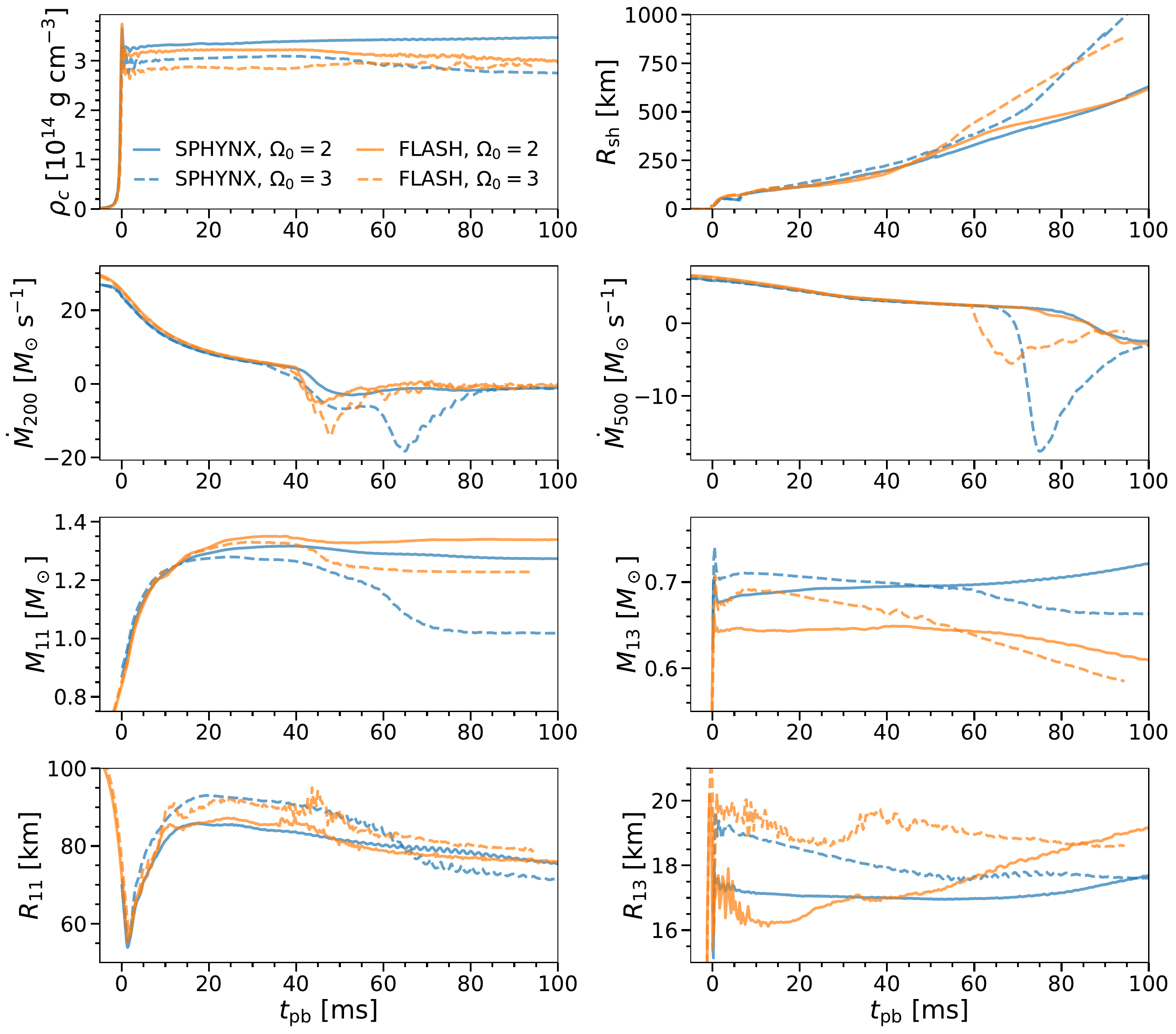}
	\caption{Panels from left to right and from top to bottom describe the time evolution of the central density ($\rho_c$), the averaged shock radius ($R_\mathrm{sh}$), the mass accretion rate measured at $r=200$~km and $500$~km ($\dot{M}_{200}$ and $\dot{M}_{500}$), the enclosed mass within a density contour of $10^{11}$~\gcm{} and $10^{13}$~\gcm{} ($M_{11}$ and $M_{13}$), and the averaged isodensity radii corresponding to $M_{11}$ and $M_{13}$ ($R_{11}$ and $R_{13}$) in the models S20H, S30H, F20, and F30.}
	\label{fig:CCSN_dynamic}
\end{figure*}

\begin{figure}
    \includegraphics[width=0.42\textwidth]{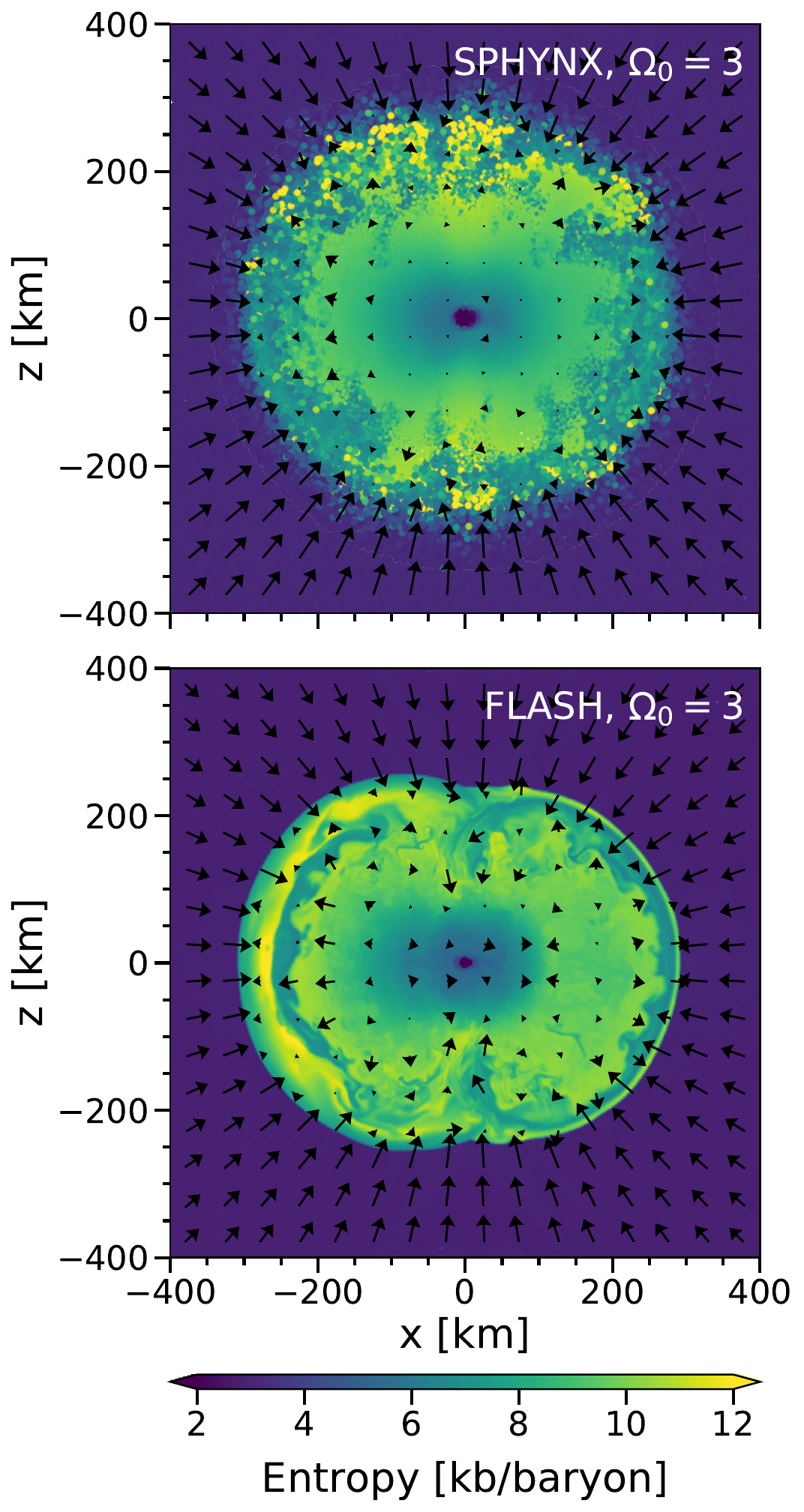}
	\caption{Entropy distribution on the XZ-plane at $t_\mathrm{pb} = 50$~ms in models S30H (top) and F30 (bottom). The arrows represent the velocity field.}
	\label{fig:slice_entr_wvelo}
\end{figure}

\begin{figure*}
    \includegraphics[width=1.0\textwidth]{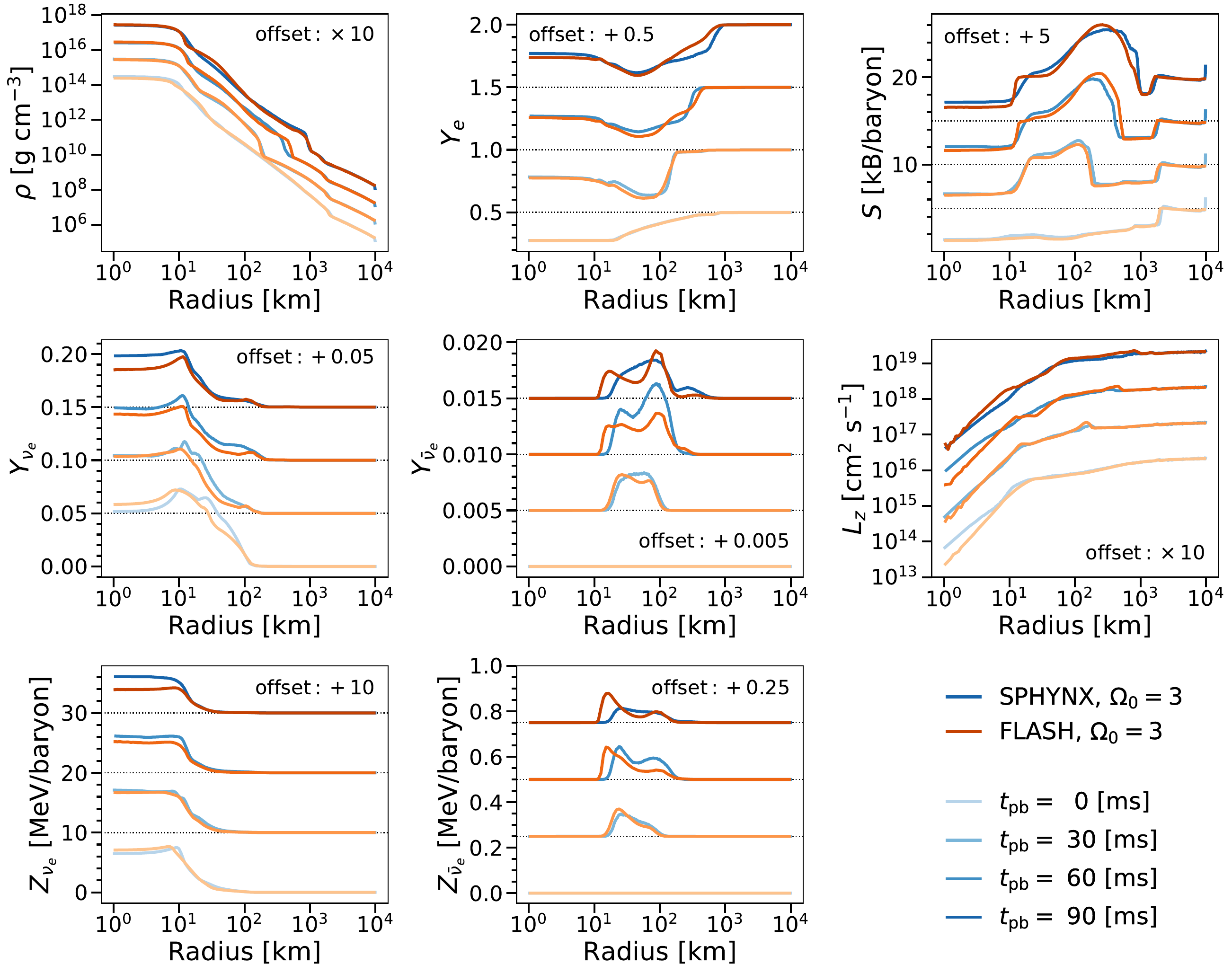}
	\caption{Spherically-averaged, radial profiles of various fluid and neutrino quantities at different postbounce times for models S30H (blue) and F30 (orange). The profiles at different times are shifted cumulatively by the offset labeled in each panel, where the black dotted lines denote the corresponding zero point.}
	\label{fig:profile_evol}
\end{figure*}

We first summarize and compare the evolution of the collapsar in our simulations performed with \sphynx{} and \flash{}. 
A broad overview of the time evolution of several PNS quantities is shown in Figure~\ref{fig:CCSN_dynamic} for models with $\Omega_0 = 2$ and $3$~\rads{} (models S20H, S30H, F20, and F30).
The panels in the first row of Figure~\ref{fig:CCSN_dynamic} show the time evolution of the central density and averaged shock radius, respectively, which are roughly in agreement between both hydrodynamics codes.
As pointed out in \cite{2018A&A...619A.118C}, \sphynx{} simulations generally produce slightly higher central densities than \flash{} simulations. This is due to the different definitions of the central density. In \flash{}, the central density is taken from the cell-averaged density of the single densest cell ($\Delta x = 488$~m), while for \sphynx{}, it is evaluated as an average of the $50$ densest SPH particles, with an equivalent cell width of less than $300$~m.
From the averaged shock radius evolution, we find that models with both initial angular velocities exhibit a very rapid explosion, where the averaged shock radii exceed $500$~km within $70-100$~ms postbounce. This is much faster than the previous work with slower initial angular velocities \citep{2016MNRAS.461L.112T, 2019MNRAS.486.2238A, 2021ApJ...914..140P, 2021MNRAS.502.3066S}, but comparable with the explosion times in \citet{2014PhRvD..89d4011K}.

In the second row of Figure~\ref{fig:CCSN_dynamic}, we show the mass accretion rates measured at $r = 200$ and $500$~km, which are in good agreement between codes in the models with $\Omega_0 = 2$~\rads{} (solid lines). The sudden drops in the mass accretion rates occur when the shock front reaches the measured radii at $t_\mathrm{pb} \sim 40$ and $80$~ms, respectively, transitioning from infall to outflow.
The $\Omega_0 = 3$ cases (dashed lines) show behavior similar to that of the $\Omega_0 = 2$ cases, though the deviations become larger when the shock front reaches the measured radii. The more substantial negative mass accretion rate in the \sphynx{} model (S30H) implies a stronger explosion compared to the \flash{} model (F30).
This interpretation can also be comprehended from the entropy distribution as well. 
Figure~\ref{fig:slice_entr_wvelo} shows slice color plots of the entropy distribution at $50$~ms postbounce along the XZ-plane, perpendicular to the equatorial plane with the $z$-axis representing the rotation axis, using models S30H and F30. Both models have similar shock expansion and convective regions, but S30H has a more spherically symmetric shock expansion at around $300$~km due to poor shock resolution ($\Delta x \sim 25$~km) at this low-density region. 
In consequence, the negative accretion rate contributed from the outflow near the equatorial plane is negated by the inflow through the pole in model F30, leading to a less negative mass accretion rate compared to model S30H. Furthermore, the drops in the mass accretion in model S30H occur later than model F30 due to the slower averaged shock expansion.

The panels in the third row of Figure~\ref{fig:CCSN_dynamic} show the evolution of the enclosed mass within a density contour of $10^{11}$~\gcm{} ($M_{11}$) and $10^{13}$~\gcm{} ($M_{13}$), respectively, and the corresponding isodensity radii ($R_{11}$ and $R_{13}$) in the last row.
The radius at $R_{11}$ usually coincides with the neutrino sphere, and therefore, the enclosed mass $M_{11}$ is used to represent the PNS mass. The enclosed mass $M_{13}$ roughly describes the PNS inner core. 
In the models with $\Omega_0 = 2$~\rads{}, model S20H has a slightly lower enclosed mass $M_{11}$ compared to model F20. 
This is mainly due to the lower mass accretion rates in \sphynx{} around core bounce, while the PNS radius $R_{11}$ remains similar. 
In the cases of $\Omega_0 = 3$~\rads{}, the deviation in the PNS mass $M_{11}$ increases further after $t_\mathrm{pb} = 40$~ms when the low-\tw{} instability starts to develop and generates stronger asymmetric accretion at later stages postbounce. 
On the other hand, the PNS inner core mass $M_{13}$ shows the opposite behavior. 
In both angular velocities, \sphynx{} models have more compact inner cores than their \flash{} counterparts due to better core resolutions in models S20H and S30H (see Table~\ref{table:models}).

Typically, when there is no rotation, the central density and the PNS inner core mass $M_{13}$ are expected to increase over time due to ongoing mass accretion and PNS cooling. 
However, in the case of rotating progenitors, we observe that model S20H has a slight increase in the inner core mass $M_{13}$ within $100$~ms postbounce, whereas model F20 shows almost no increase within the first $40$~ms postbounce and is followed by a slight decrease in the inner core mass. These distinctions may be due to differences in angular momentum transport and conservation between \sphynx{} and \flash{} as described in \cite{2018A&A...619A.118C}. 
In the models with $\Omega_0 = 3$~\rads{}, when angular momentum keeps propagating inward, the centrifugal force eventually overcomes the gravitational force and, therefore, it induces a decrease in the PNS inner core mass $M_{13}$. Both models S30H and F30 behave similarly but with different decreasing rates.  

In addition, we also find that \flash{} simulations have a notable neutron star kick and a modulation motion ($v_\mathrm{pns} \sim 240$ and $360$~\kms{} in models F20 and F30, respectively) due to an asymmetric explosion and together with numerical artifacts on angular momentum non-conservation. These motions affect the evolution of the PNS in \flash{} simulations, especially at late time. On the other hand, the neutron star kick velocities in S20H and S30H are less than $1$~\kms{}. In the following sections, we show that these differences in the compactness and motion of the PNS inner core will affect the GW signatures of the low-\tw{} instability.

A comparison of radial profiles of various fluid and neutrino quantities between the models S30H and F30 at different postbounce times is shown in Figure~\ref{fig:profile_evol}.
The spherically-averaged profiles of density, electron fraction, and entropy are consistent in the radius coordinate.
The reason for the lower angular momentum observed in the inner core of the PNS in model F30 is attributed to the fact that angular momentum conservation has deteriorated, which is primarily caused by inherent numerical dissipation in grid-based hydrodynamics codes (as discussed in \citealt{2018A&A...619A.118C}).
When considering the quantities of electron-type neutrinos, it should be noted that model S30H has a more compact and hotter inner core in its PNS in comparison to the corresponding PNS in model F30 (see Figure~\ref{fig:CCSN_dynamic}). This leads to a higher production rate and energy level of electron neutrinos within the PNS inner core in model S30H. Overall, the radial profiles of neutrino quantities at different times are consistent between the models S30H and F30.

\begin{figure*}
	\plotone{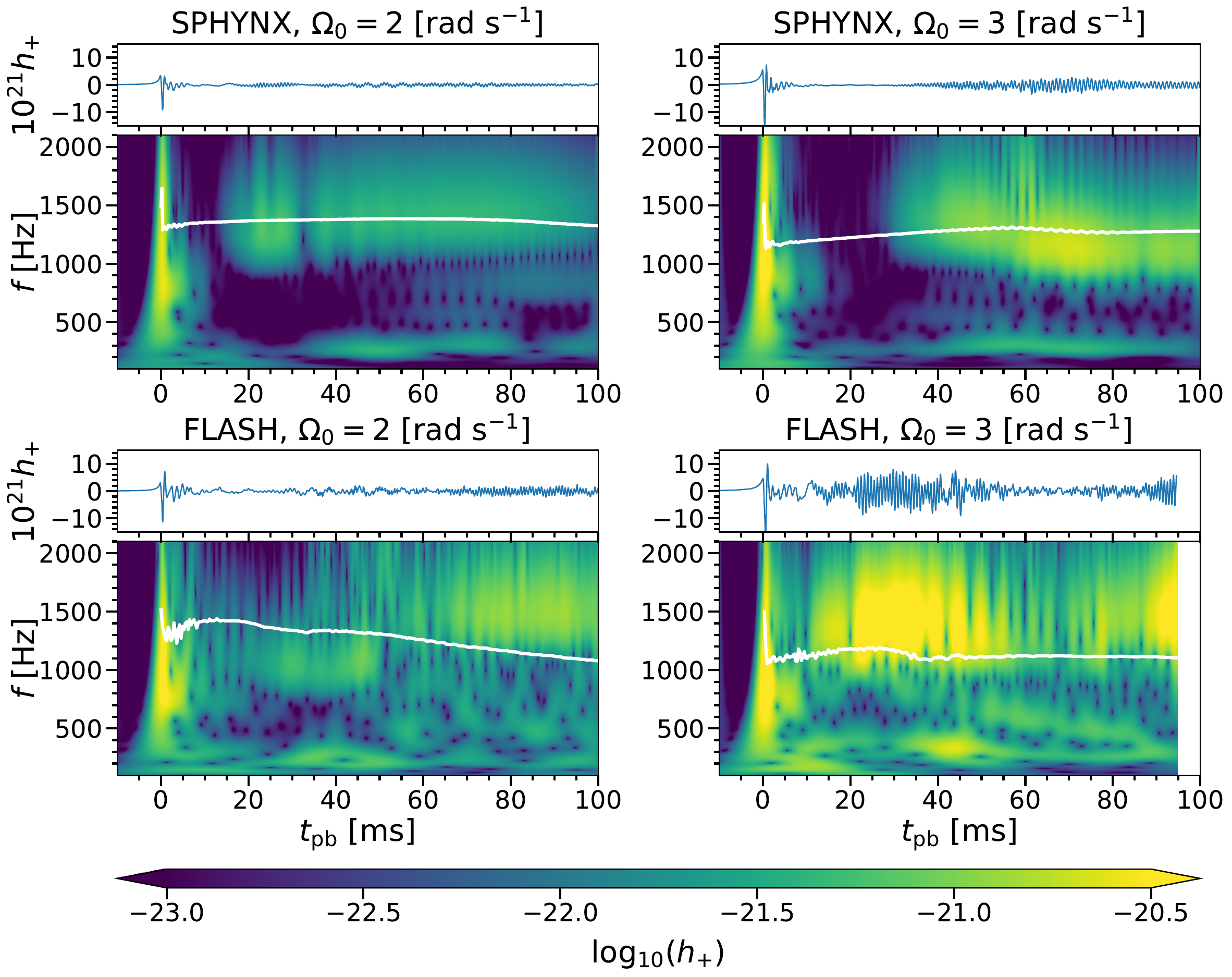}
	\caption{GW strains and spectrograms of the plus mode for the models with $\Omega_0 = 2$ and $3$~\rads{}, using \sphynx{} with $1600$k particles (top panels) and \flash{} (bottom panels), seen along the equatorial plane at a distance of $10$~kpc. The white line represents the doubled dynamical frequency at a density of $10^{13}$~\gcm{} (see Section~\ref{sec:gw_omega} for a more detailed description).}
	\label{fig:GW_both}
\end{figure*}

\subsection{GW Features and Origin in Rapidly Rotating CCSNe}  \label{sec:gw_code}

In this section, we discuss the GW features in the models with $\Omega_0 = 2$ and $3$~\rads{}, using \sphynx{} with $1600$k particles (models S20H and S30H) and \flash{} (models F20 and F30). 
The top panels in Figure~\ref{fig:GW_both} show the plus mode of GW emissions from models S20H and S30H, seen along the equatorial plane at a distance of $10$~kpc. 
The GW strain is shown at the top, and the corresponding spectrogram is displayed at the bottom in each panel. The spectrogram is computed using the wavelet analysis implemented in the PyCWT\footnote{\url{https://github.com/regeirk/pycwt}} code, a Python package based on \citet{1998BAMS...79...61T}, where the power spectrum is divided by the wavelet scales to rectify the energy bias \citep{2007JAtOT..24.2093L}. 
In addition, we divide the resulting amplitude by the square root of the sampling rate to ensure consistent strength between different sampling rates.

\begin{figure*}
    \includegraphics[width=0.95\textwidth]{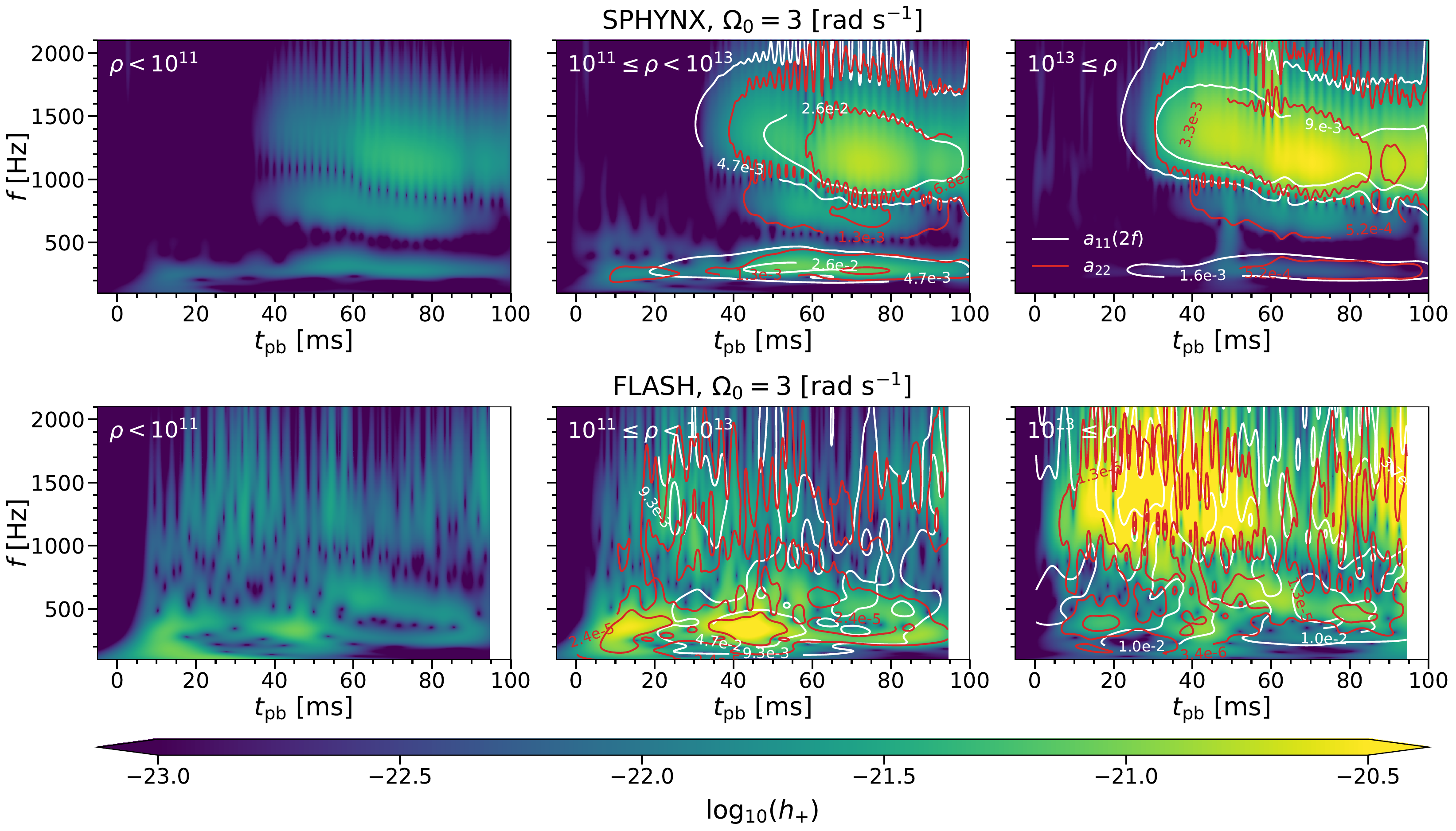}
	\caption{GW spectrograms of the plus mode emitted from regions with density $\rho < 10^{11}$~\gcm{} (left), $10^{11} \leq \rho < 10^{13}$~\gcm{} (middle), and $\rho \ge 10^{13}$~\gcm{} (right) for models S30H (top) and F30 (bottom), seen along the pole at a source distance of $10$~kpc. In the middle and right panels, the white and red contours show the spectrogram of the normalized spherical harmonic coefficients $a_{11}$ and $a_{22}$, respectively, where the frequency of $a_{11}$ is doubled to ease comparison.}
	\label{fig:GW_SphHarm_both}
\end{figure*}

First, both \sphynx{} models show distinctive bounce and ring-down signals in the first $20$~ms postbounce, which have been extensively studied in previous works \citep[e.g.,][and references therein]{2014PhRvD..90d4001A, 2017PhRvD..95f3019R, 2022hgwa.bookE..21A}.
After the bounce and ring-down signals, we can observe that the GW emissions exhibit quasi-sinusoidal time oscillations starting at $t_\mathrm{pb} = 20 - 30$~ms in both models. 
We find that this GW feature is simultaneous with the so-called low-\tw{} instability
\citep{2003ApJ...595..352S, 2005ApJ...625L.119O, 2007PhRvL..98z1101O, 2008A&A...490..231S, 2010A&A...514A..51S, 2014PhRvD..89d4011K, 2020MNRAS.493L.138S}. 
The low-\tw{} instability is a non-axisymmetric rotational instability that develops in cores with a high degree of differential rotation around the corotation radius, where the pattern frequency of the induced oscillation is equal to the local angular frequency of the background flow \citep{2001ApJ...550L.193C, 2005ApJ...618L..37W, 2006MNRAS.368.1429S}. 
This low-\tw{} instability can induce $m = 1$ and/or $m = 2$ deformations that lead to a time-changing quadrupole moment, which is the ultimate source of the GW emissions. 

From the spectrograms of S20H and S30H in Figure~\ref{fig:GW_both}, we can identify two strong GW signals in the frequency ranges from $200$ to $400$~Hz and from $1100$ to $1400$~Hz. Hereafter we refer to them as the $300$~Hz signal and the kHz signal, respectively.
In the bottom panels of Figure~\ref{fig:GW_both}, we present the GW strains obtained using the \flash{} code with $\Omega_0 = 2$~\rads{} (F20) and $\Omega_0 = 3$~\rads{} (F30), and their corresponding spectrograms. The GW emissions in both models exhibit similar bounce, ring-down, $300$~Hz, and kHz signals as discussed above for the \sphynx{} simulations. Although the bounce and ring-down signals in models F20 and F30 are consistent with the S20H and S30H models, we can see that the $300$~Hz and kHz signals evolve differently, especially with respect to the occurrence time and peak frequency of the kHz signal. 
As discussed in Section~\ref{sec:dynamic_overview}, \flash{} and \sphynx{} show slightly different dynamical evolutions of the PNS after core bounce (see Figure~\ref{fig:CCSN_dynamic}). In Section~\ref{sec:gw_omega}, we will show that the $300$~Hz and kHz signals are correlated with the outer and inner structures of the PNS, respectively, and therefore causes the differences between \flash{} and \sphynx{} models.

To investigate the origin of the $300$~Hz and kHz signals, we calculate the contributions of GW emissions from different density regions by post-processing the simulation data, using the formulae described in Section~\ref{sec:method_gw}. 
Figure~\ref{fig:GW_SphHarm_both} shows the GW contributions from regions within density $\rho < 10^{11}$, $10^{11} \le \rho < 10^{13}$, and $\rho \ge 10^{13}$~\gcm{} for models S30H and F30, seen along the pole to eliminate bounce and ring-down signals. In both models, we can see that the $300$~Hz and kHz signals emanate from separate regions. 
The $300$~Hz signal is mainly from the region of $10^{11} \le \rho < 10^{13}$~\gcm{}, while the kHz signal is mainly from the PNS inner core where $\rho \ge 10^{13}$~\gcm{}.

\begin{figure*}
	\plotone{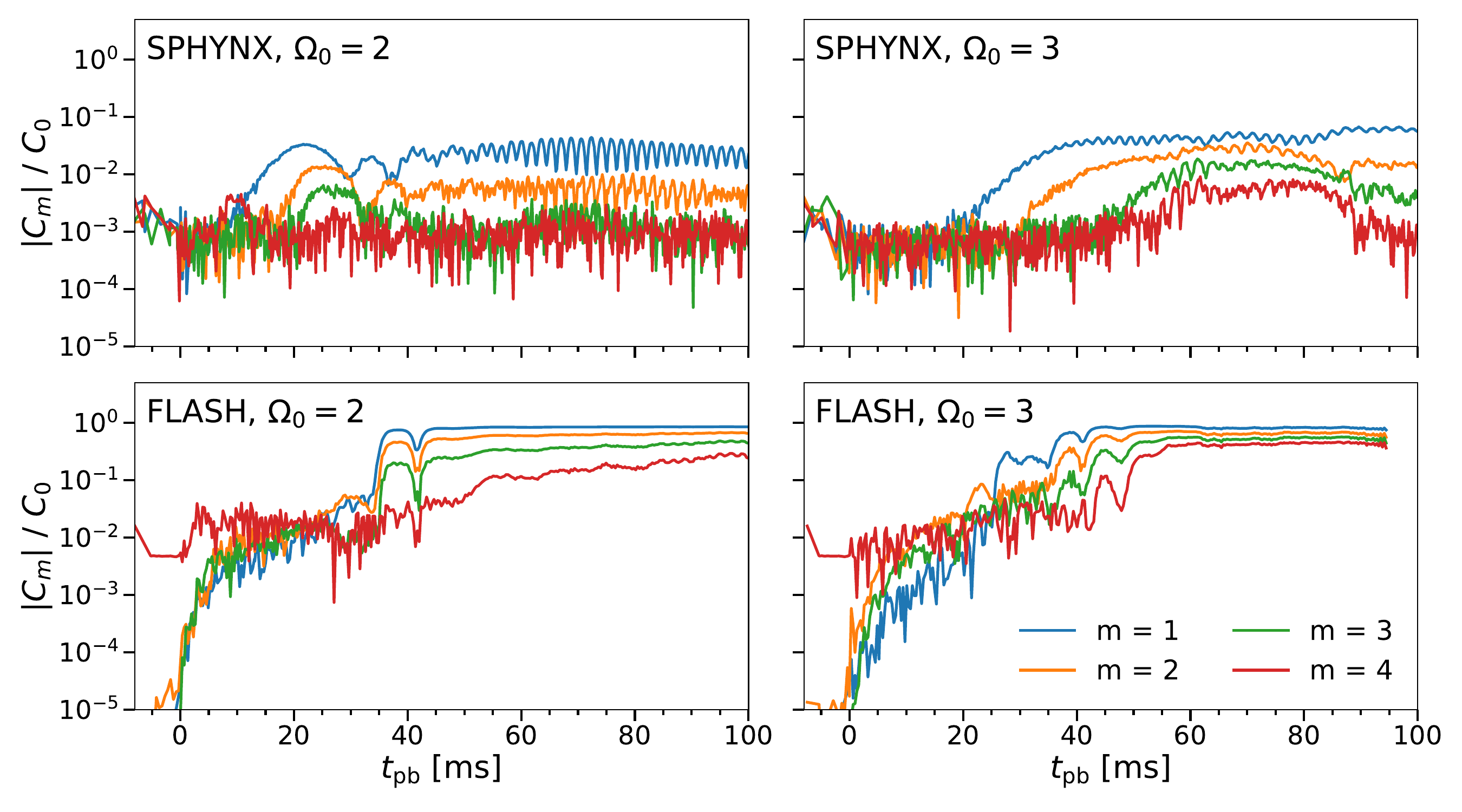}
	\caption{Normalized mode amplitude at a radius of $15$~km for models S20H and S30H (top), and F20 and F30 (bottom).}
	\label{fig:imphi}
\end{figure*}

In Figure~\ref{fig:GW_SphHarm_both}, we also evaluate the spherical harmonic components $a_{11}$ and $a_{22}$ in the regions of $10^{11} \leq \rho < 10^{13}$ and $\rho \geq 10^{13}$~\gcm{} (defined in Section~\ref{sec:method_sphharm}), and overlap their spectrograms in white and red lines, respectively.
The components $a_{11}$ and $a_{22}$ represent the $m = 1$ and $m = 2$ deformations in the specific density region, and their contours indicate the mode frequency of the corresponding deformation, $f_{\mathrm{mode}, m}$. The mode frequency is related to the pattern frequency by $f_{\mathrm{pat}, m}= f_{\mathrm{mode}, m} / m$ \citep{2005ApJ...618L..37W}. 
Note that the mode frequency of $a_{11}$ is doubled in Figure~\ref{fig:GW_SphHarm_both} to facilitate comparison between the components $a_{11}$ and $a_{22}$.
Comparing the spherical harmonic modes with the $300$~Hz and kHz GW signals reveals that the $a_{22}$ component coincides with both. This is because the dominant quadrupole component of GW emissions stems from the $l = 2, m = 2$ mode, making it a natural source for both signals.
On the other hand, the pattern frequencies of $a_{11}$ and $a_{22}$ satisfy the relation $f_{\mathrm{pat}, 1} \simeq f_{\mathrm{pat}, 2}$, indicating that the $a_{22}$ component is a daughter mode of the $a_{11}$ component.
This infers that both the $300$~Hz and kHz signals are associated with the $m = 1$ spiral deformation induced by the low-\tw{} instability.
\citet{2020MNRAS.493L.138S} conducted a full-GR CCSN simulation of a rapid-rotating $70 M_\odot$ progenitor. They found a transient quasi-periodic time modulation at $450$~Hz from the $m = 1$ spiral deformation in $50-100$~km. 
The $300$~Hz signal in our models is similar to the $450$~Hz signal in \citet{2020MNRAS.493L.138S}.
On the other hand, the kHz signal resembles the $\sim 930$~Hz signal found by \citet{2007PhRvL..98z1101O} in CCSN simulations of a $20 M_\odot$ progenitor, which is correlated with the $m = 1$ mode at $10-15$~km, but the peak frequency is higher in our cases.
In addition to the $300$~Hz and kHz signals, some higher-order modes of GW emissions at around $800$~Hz can be seen in Figure~\ref{fig:GW_SphHarm_both} as well. In model S30H, the $800$~Hz signal is correlated to the kHz signal and the $a_{22}$ component but is much weaker than the $300$~Hz and kHz signals. In model F30, similar higher-order modes also exist between the $300$~Hz and kHz signals, but the interactions among these GW features are more complex than that in model S30H.

\begin{figure*}
	\plotone{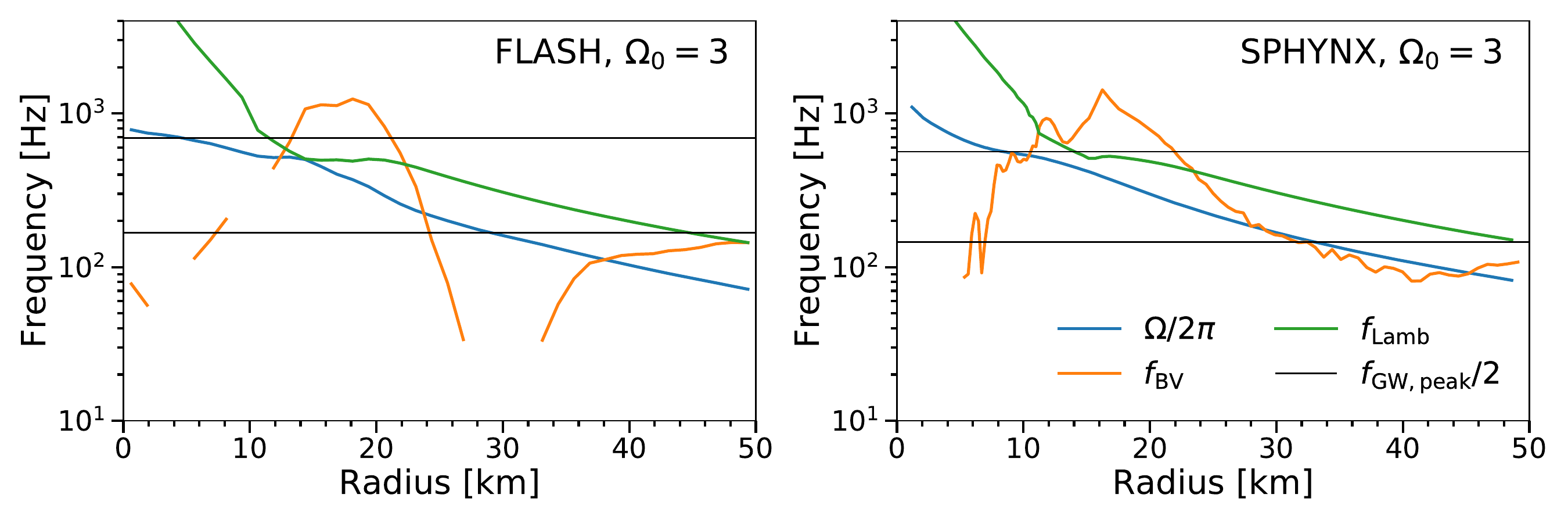}
	\caption{Radial profiles of rotational (blue), \bvfreq{} (orange), and Lamb (green) frequencies at $t_\mathrm{pb} = 30$~ms for models F30 (left) and S30H (right). The black solid lines denote the half-peak frequencies of the $300$~Hz and kHz gravitational-wave signals.} 
	\label{fig:BV_Lamb}
\end{figure*}

\begin{figure*}
    \includegraphics[width=1.0\textwidth]{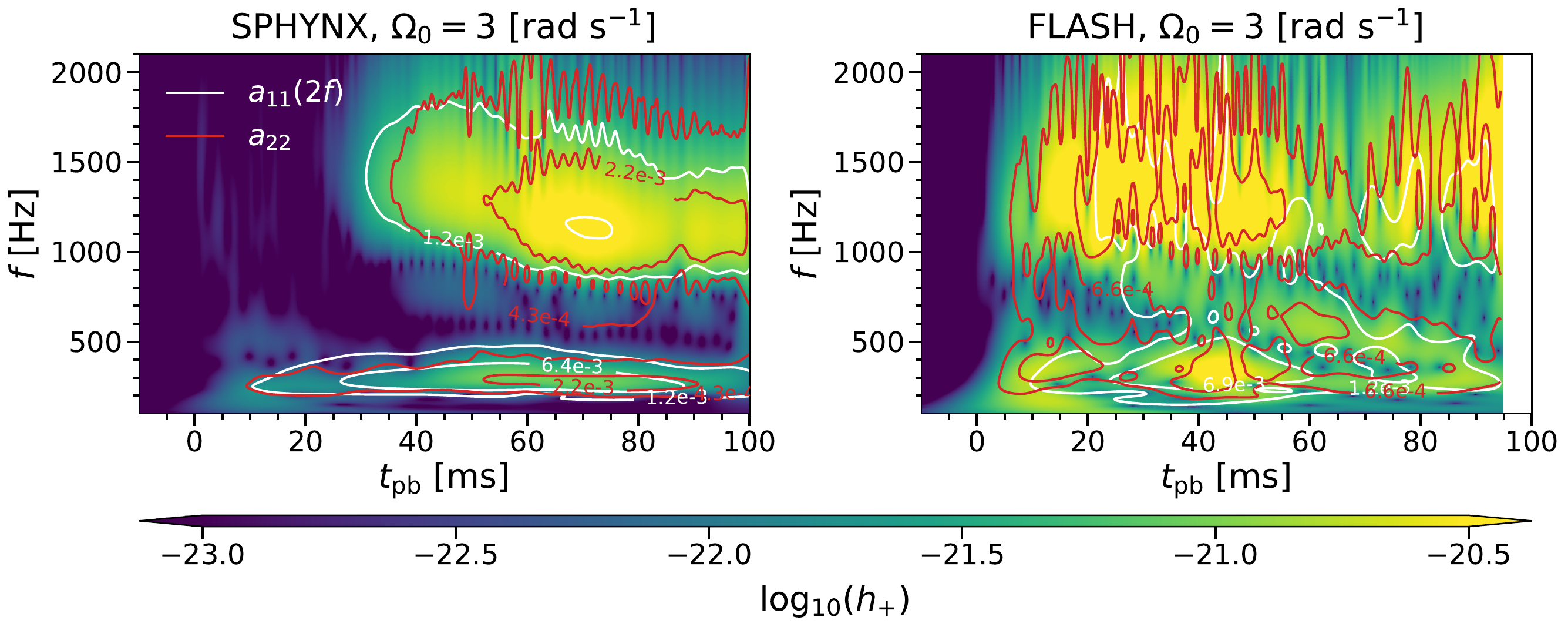}
	\caption{Spectrograms of the normalized spherical harmonic coefficients, $a_{11}$ (white, doubled frequency) and $a_{22}$ (red), of the mean energy of trapped electron anti-neutrinos for models S30H (left) and F30 (right). The GW spectrograms of the plus mode, seen along the pole at $10$~kpc, are shown in filled contour to ease comparison.}
	\label{fig:SphHarm_Enua_both}
\end{figure*}

To investigate the potential impact of $m = 4$ perturbations induced by the Cartesian grid discretization in \flash{} simulations,
we also perform an analysis of the azimuthal density modes in the equatorial plane by computing the Fourier amplitude \citep{2001ApJ...550L.193C}
\begin{equation} \label{eq:c_m}
C_m(\varpi) = \frac{1}{2\pi} \int^{2\pi}_0 \rho(\varpi, z = 0) e^{im\phi} d\phi,
\end{equation}
where $\varpi$ is the cylindrical radius relative to the PNS center. 
Figure~\ref{fig:imphi} shows the normalized mode amplitude, $|C_m| / C_0$, evaluated at a radius of $\varpi = 15$~km. 
We note that the relative difference in the mean density, $C_0$, is below $30\%$ before $t_\mathrm{pb} \sim 35$~ms between \sphynx{} and \flash{} simulations and then increases to $80\%$ due to the PNS kicks in \flash{} simulations, making the normalized mode amplitudes in \flash{} simulations higher than the \sphynx{} counterpart simulations. 
Furthermore, we can see that in \flash{} simulations, the $m = 4$ mode is the dominant mode in the early postbounce phase ($t_\mathrm{pb} \lesssim 10$~ms) due to grid effects, while it is always subdominant in \sphynx{} simulations, as expected for a meshless method. 
However, there is no clear relation between the $m = 4$ grid mode and the other $m = \{1, 2, 3\}$ modes in \flash{} simulations, and thus both F20 and F30 models remain dynamically stable to grid perturbations. This is also consistent with the earlier work done with the full GR simulations via the Whisky code in \citet{2007PhRvL..98z1101O}. 

Recently, \citet{2021MNRAS.508..966T} proposed that the low-\tw{} instability in CCSN environments could be triggered by the Rossby waves growing near the convective zone.
In their explanation, such instability requires having a corotation radius to coincide with the convective layer in the PNS. 
Therefore, it is interesting to investigate whether the $300$~Hz and kHz low-\tw{} signals in our models are satisfied with the same criteria. Figure~\ref{fig:BV_Lamb} shows the radial profiles of the rotational, \bvfreq{}, and Lamb frequencies, evaluated in the equatorial plane, for models F30 and S30H at $t_\mathrm{pb} = 30$~ms. We evaluate the Lamb frequency via
\begin{equation}
f_\mathrm{Lamb} = \frac{1}{2 \pi} \frac{\sqrt{l (l + 1)} c_s}{r},
\end{equation}
and the \bvfreq{} frequency using the Ledoux criterion \citep{2006A&A...447.1049B, 2013ApJ...768..115O}
\begin{equation}
f_\mathrm{BV} = \frac{\mathrm{sign}(C_L)}{2 \pi} \sqrt{\left| \frac{C_L}{\rho} \frac{d\Phi}{dr} \right|},
\end{equation}
where $l$ is taken to be $1$, $c_s$ is the local speed of sound, $d\Phi$ is the local gravitational potential where the approximation $d\Phi /dr \sim -G M(r) / r^{2}$ is adopted. The Ledoux criterion reads \citep{1947ApJ...105..305L}
\begin{equation}
C_L = -\left( \frac{\partial \rho}{\partial P} \right)_{s, Y_l}
      \left[ \left( \frac{\partial P}{\partial s}  \right)_{\rho, Y_l} \left( \frac{ds}{dr}   \right)
           + \left( \frac{\partial P}{\partial Y_l} \right)_{\rho, s}  \left( \frac{dY_l}{dr} \right)\right],
\end{equation}
where we approximate the lepton fraction by the electron fraction, $Y_l \sim Y_e$, for simplicity. 
In model F30, two corotation radii at $5-10$~km and $30$~km, which are described by the intersection of the pattern frequencies of the $300$~Hz and kHz signals (black lines) and the rotational frequency (blue line), are in convective regions with negative \bvfreq{} frequency. This is consistent with the statement proposed by \citet{2021MNRAS.508..966T}.
On the other hand, in the case of model S30H, only one corotation radius of the kHz signal at $5-10$~km coincides with the edge of a convective layer. However, we note that the \bvfreq{} frequency oscillates rapidly between $30$ and $40$~km in S30H, suggesting that a convective zone could be developing in that region as well.

In addition to the Rossby wave scenario, it is worth mentioning that the area where the kHz signal originated, approximately at around the isodensity radius $R_{13}$ ($r \sim 20$~km), aligns with the region where electron anti-neutrinos are predominantly generated and remain coupled with the matter.
Therefore, the density variation driven by the $m = 1$ deformation can affect the production of electron anti-neutrinos in the PNS inner core.
In the IDSA neutrino treatment, electron-type neutrinos are decomposed into trapped and free-streaming neutrinos. Among those, only trapped neutrinos are coupled with the fluid.  
Since in our implementation of IDSA, the $\mathcal{O}(v/c)$ terms of the neutrino pressure are included (see Equation~24 in \citealt{2009ApJ...698.1174L}), the neutrino pressure gradient could contribute to the development the $m = 1$ deformation. 
To establish this, we apply the spherical harmonic decomposition to the mean energy of trapped electron anti-neutrinos through a variant of Equation~(\ref{eq:sphharm}):
\begin{equation}
a_{lm} = \frac{ (-1)^{|m|} }{ \sqrt{4\pi(2l + 1)} } 
         \frac{\sum w \left< E_{\bar{\nu}_e} \right>(\theta,\phi) Y_l^m(\theta,\phi)}
              {\sum w},
\end{equation}
where $\left< E_{\bar{\nu}_e} \right> = Z_{\bar{\nu}_e} / Y_{\bar{\nu}_e}$ is the mean energy of trapped electron anti-neutrinos.
We use the mean energy $\left< E_{\bar{\nu}_e} \right>$, instead of $Y_{\bar{\nu}_e}$ or $Z_{\bar{\nu}_e}$, to avoid the sharp decline in electron anti-neutrino fractions, which could introduce strong numerical noise.  
We evaluate the spherical harmonic components in the shell $R_{13} \pm 10$~km.
Figure~\ref{fig:SphHarm_Enua_both} shows the spectrograms of the $a_{11}$ and $a_{22}$ components for models S30H and F30, revealing that the distributions of trapped electron anti-neutrinos are in good agreement with not only the kHz signal but also the $300$~Hz signal.
Albeit the asymmetric electron anti-neutrino distribution requires the $m = 1$ deformation from the low-\tw{} instability to grow. 
In the following section, we show that there are distinct phase differences between the $m = 1$ deformation and the asymmetric neutrino distribution in the PNS inner core.
This suggests that neutrino pressure could play a role in promoting the development of $m = 1$ deformation and exhibiting unique GW signatures in the kHz window.
We also find that the distribution of electron neutrinos has a similar asymmetric effect but is less pronounced than electron anti-neutrinos. This is because electron neutrinos have been produced since the collapse.

\subsection{GW Dependence on Initial Angular Velocity} \label{sec:gw_omega}

\begin{figure}
    \plotone{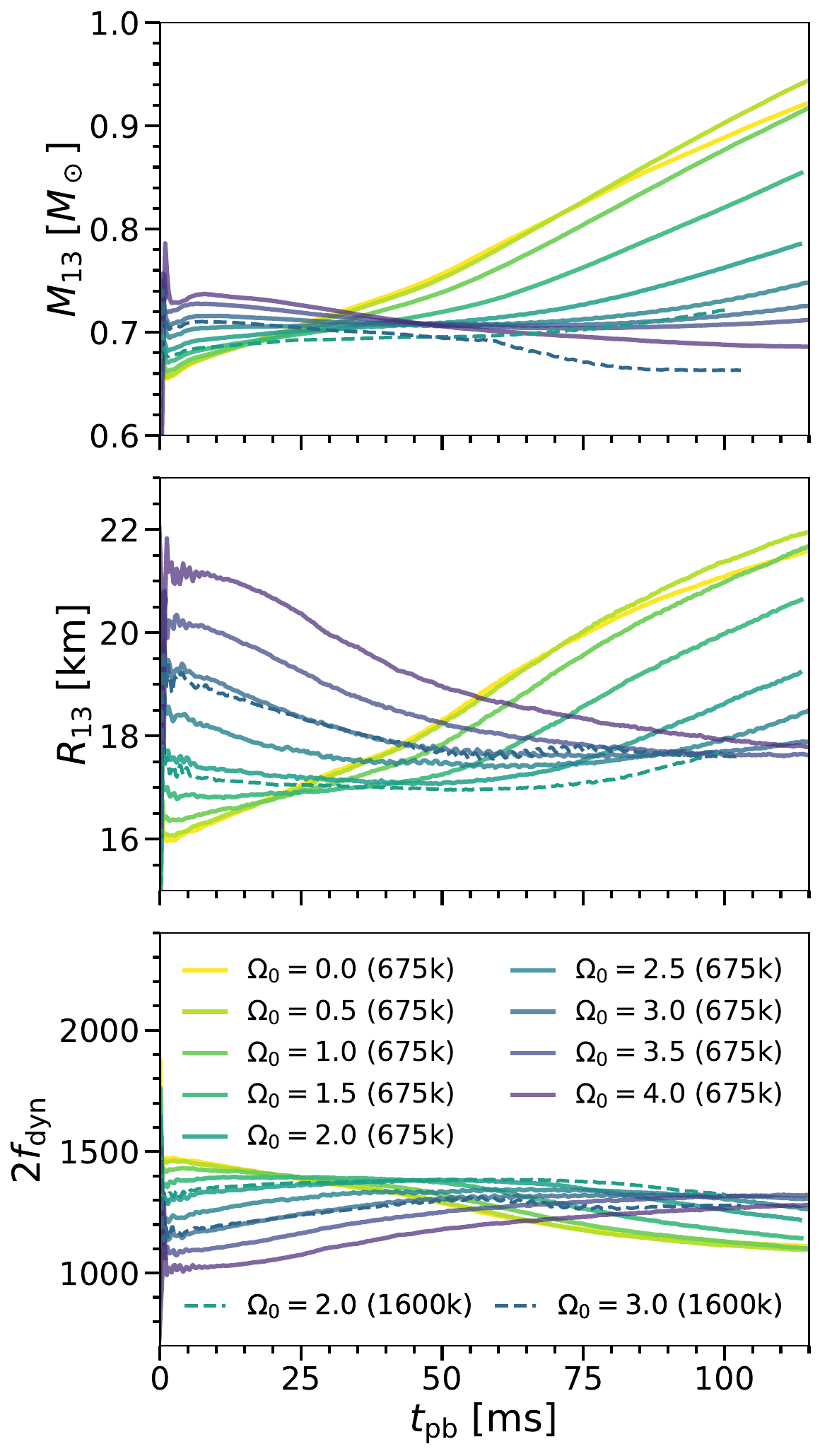}
    \caption{Time evolution of the PNS mass with density \mbox{$\rho \geq 10^{13}$~\gcm{}} (top), the corresponding averaged isodensity radius (middle), and the doubled dynamical frequency (bottom) in the \sphynx{} simulations with $675$k particles (solid lines). Different colors represent simulations with different initial angular velocities. Dashed lines show the counterpart models (S20H and S30H) with $1600$k particles for comparison.}
	\label{fig:CCSN_dynamic_675k}
\end{figure}

\begin{figure*}
    \plotone{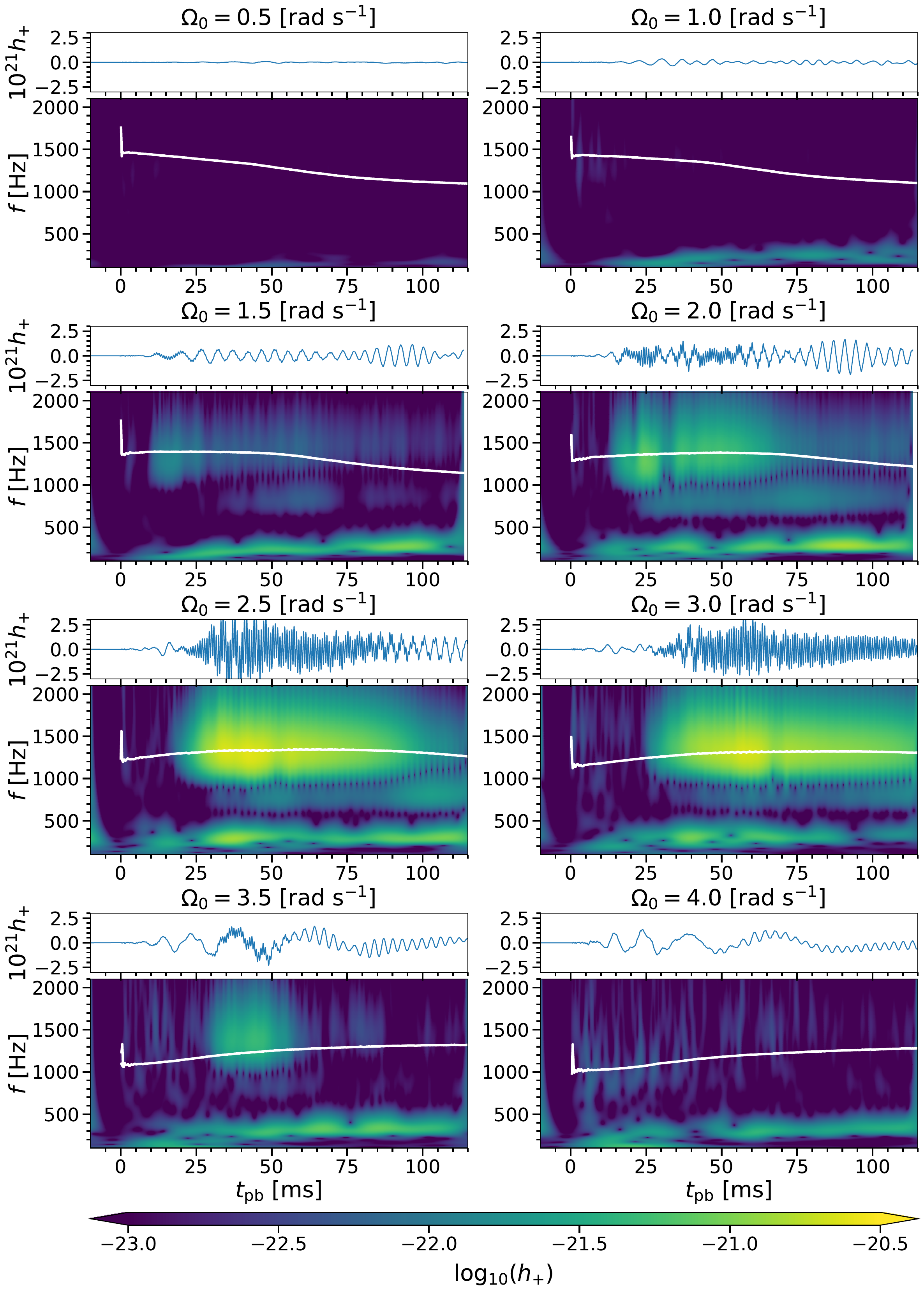}
	\caption{GW strains and spectrograms of the plus mode seen along the pole at a distance of $10$~kpc. Different panels represent the \sphynx{} $675$k particle simulations with different initial angular velocity. The white lines represent the doubled dynamical frequency for the domain with $\rho \ge 10^{13}$~\gcm{} in each model.}
	\label{fig:GW_sphynx_675k}
\end{figure*}

To investigate the dependence of the $300$~Hz and kHz low-\tw{} signals on the initial angular velocity, we perform a series of \sphynx{} simulations with $675$k particles (See Table~\ref{table:models}). 
Figure~\ref{fig:CCSN_dynamic_675k} shows the time evolution of the enclosed mass ($M_{13}$), the corresponding isodensity radius ($R_{13}$), and the doubled dynamical frequency, $2 f_{\rm dyn} \sim 2 \sqrt{R_{13}^3/GM_{13}}/2\pi$, for the domain with $\rho \ge 10^{13}$~\gcm{} for models with $\Omega_0$ ranging from $0.0$ to $4.0$~\rads{}, in steps of $0.5$~\rads{}. 
In addition, we also plot the same quantities for models S20H and S30H in dashed lines for comparison.
First, we can see that the PNS inner core structures are consistent with different particle numbers, and the trend basically follows the description provided in Section~\ref{sec:dynamic_overview} for models with $\Omega_0 = 2$ and $3$~\rads{}: higher initial angular velocities will result in a decrease of the enclosed mass and isodensity radius, while the doubled dynamical frequency remains a relatively unchanged range of kHz along the simulations.

Figure~\ref{fig:GW_sphynx_675k} shows the corresponding GW strain and spectrogram for these models, assuming viewing from the pole and at a distance of $10$~kpc. 
We note that in Figure~\ref{fig:GW_sphynx_675k}, the color ranges are fixed at the same values for all panels to facilitate comparison between different initial angular velocities. 
We first focus on the models S20 and S30 to ensure that our simulations with $675$k particles accurately capture the main GW features observed in our simulations with $1600$k particles (models S20H and S30H). 
By comparing spectrograms in Figure~\ref{fig:GW_both} and Figure~\ref{fig:GW_sphynx_675k}, we find that both the $300$~Hz and kHz signals in models S20 and S30 are similar to those of models S20H and S30H.
Since the main low-\tw{} features discussed in our high-resolution runs are also captured in the simulations with $675$k particles, we can conclude that we are able to conduct a parameter study of the initial angular velocity at a considerably lower computational cost using this resolution.

\begin{figure}
    \includegraphics[width=0.46\textwidth]{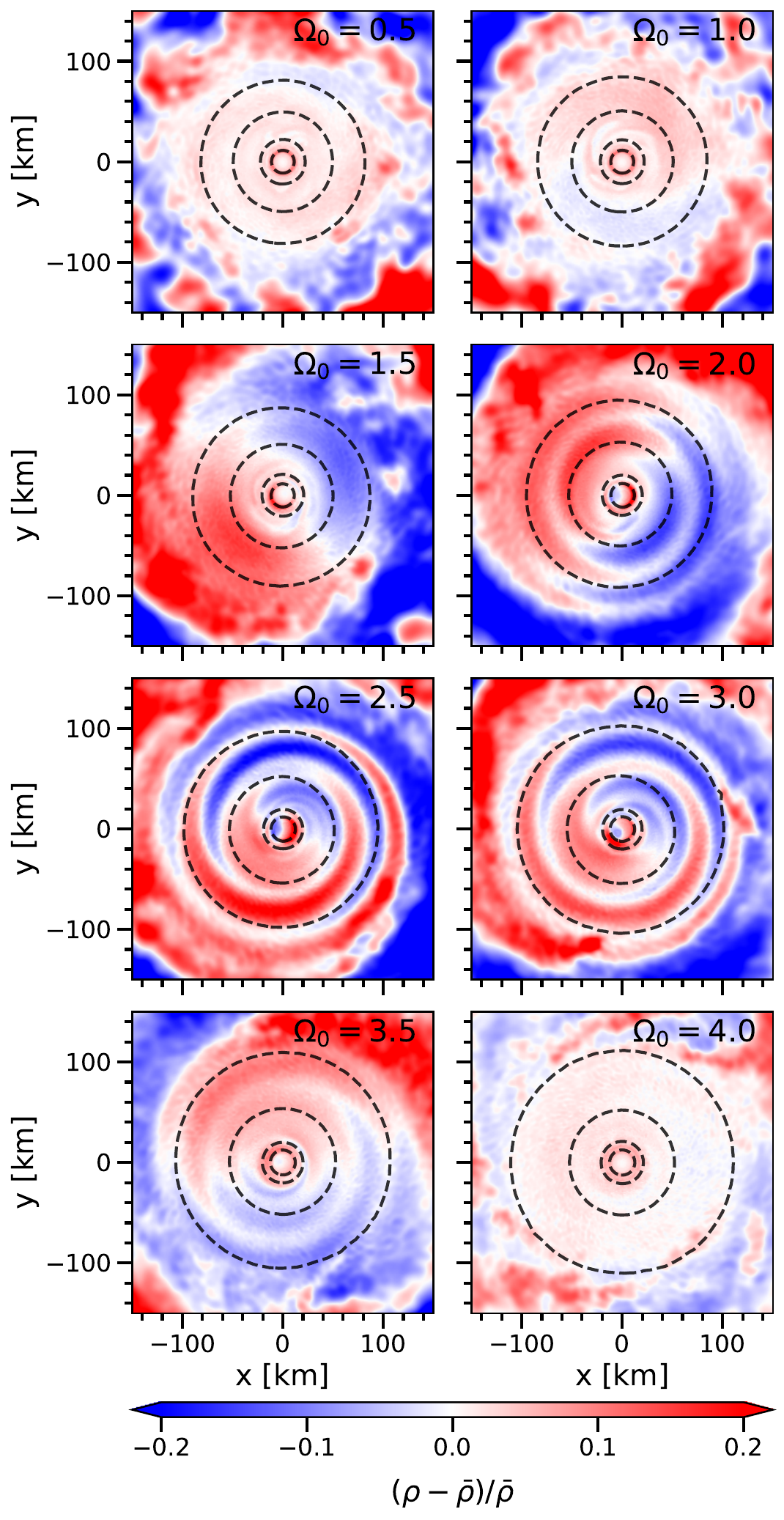}
	\caption{Normalized density fluctuation in the equatorial plane at $t_\mathrm{pb} = 50$~ms for models using \sphynx{} with $675$k particles and different initial angular velocities. The average density, $\bar{\rho}$, is taken over the azimuthal direction in the plane. The black dashed curves denote the isodensity contours at $\rho = 10^{11}$, $10^{12}$, $10^{13}$, and $10^{14}$~\gcm{}.}
	\label{fig:slice_dens}
\end{figure}

\begin{figure}
    \includegraphics[width=0.46\textwidth]{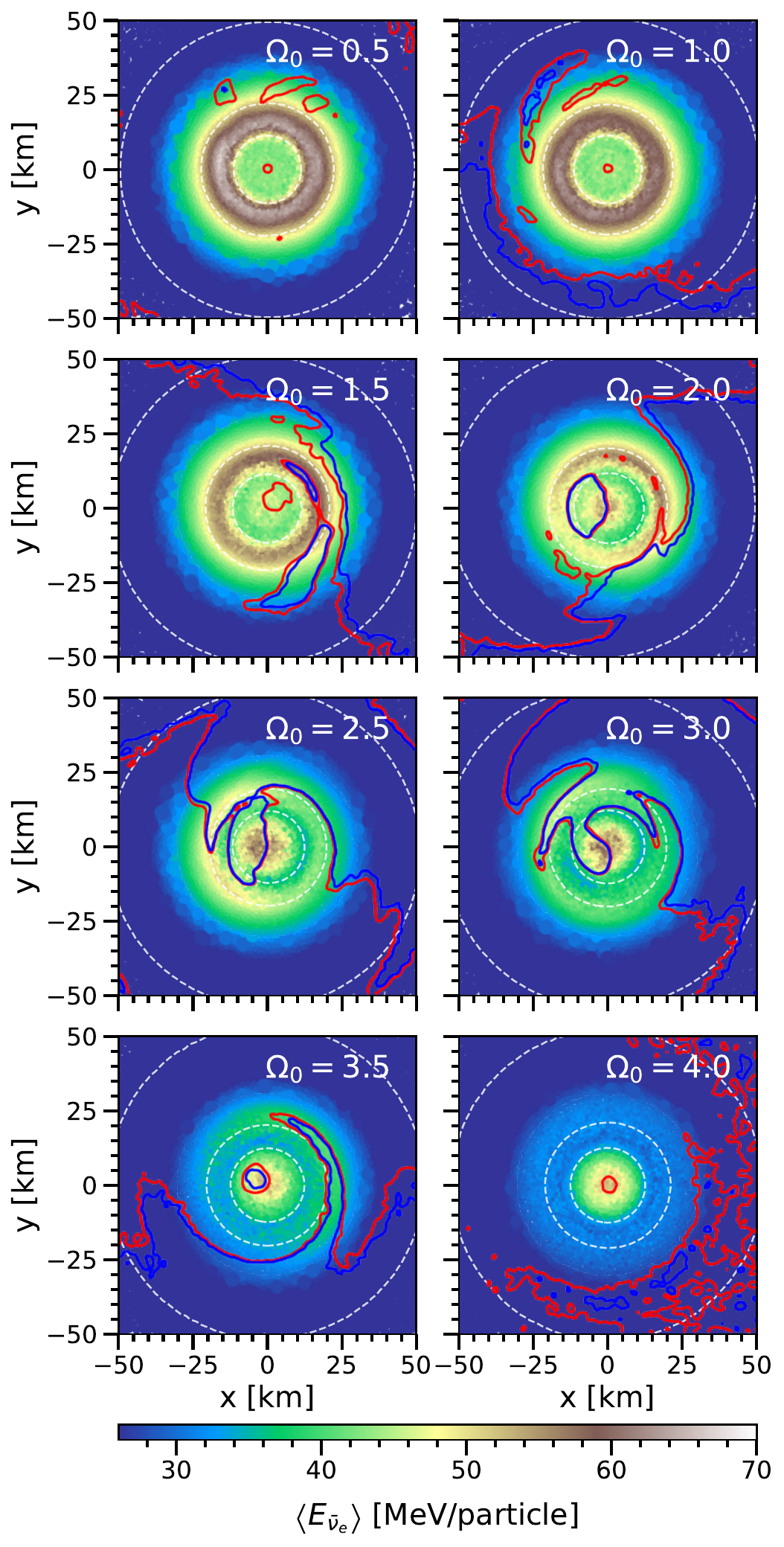}
	\caption{Mean energy of trapped electron anti-neutrino distribution on the XZ-plane at $t_\mathrm{pb} = 50$~ms for models using \sphynx{} with $675$k particles and different initial angular velocities. The red and blue curves respectively illustrate positive and negative density fluctuations centered around zero. The white dashed curves denote the isodensity contours at $\rho = 10^{12}$, $10^{13}$, and $10^{14}$~\gcm{}.}
	\label{fig:slice_Ynua}
\end{figure}

As a result, from Figure~\ref{fig:GW_sphynx_675k} we find that the $300$~Hz signal starts to appear when the initial angular velocity $\Omega_0 \ge 1$~\rads{} and it persists consistently across different angular velocities. However, the kHz signal appears only in a range within $1.5 \le \Omega_0 \le 3.5$~\rads{}, though the kHz signal in model S35 occurred in a short duration between $30-60$~ms postbounce.
In addition, we overlay the doubled dynamical frequency on the spectrograms in Figure~\ref{fig:GW_sphynx_675k}. 
We find that the kHz signal, once it appears, has a peak frequency described by the doubled dynamical frequency at the density of $10^{13}$~\gcm{}.
As we have discussed in Section~\ref{sec:gw_code}, the $m = 1$ deformation associated with the kHz signal originates from the PNS inner core where $\rho > 10^{13}$~\gcm{}, and its pattern frequency ($a_{11}$ in Figure~\ref{fig:GW_SphHarm_both}) should be related to the characteristic frequency, the dynamical frequency here, in this density region. 
Therefore, it is not surprising that the kHz signal can be described by the doubled dynamical frequency at the density of $10^{13}$~\gcm{}.

To visualize the $m = 1$ deformation, we follow \citet{2016MNRAS.461L.112T} to show the density fluctuation in the equatorial plane, $(\rho - \bar{\rho}) / \bar{\rho}$, at $t_\mathrm{pb} = 50$~ms in Figure~\ref{fig:slice_dens}, where $\bar{\rho}$ is the azimuthally averaged density.
Different isodensity contours are plotted as dashed lines as well. 
We can see that the $m = 1$ deformation can develop in several density regions.
In the density region of $10^{11} \lesssim \rho \lesssim 10^{13}$~\gcm{}, all models with $\Omega_0 \ge 1.0$~\rads{} show the spiral structure in the density fluctuation plot, but not in models with $\Omega_0 \le 0.5$~\rads{}.
This is consistent with the presence of the $300$~Hz signal in Figure~\ref{fig:GW_sphynx_675k} and also confirms that the $300$~Hz signal is indeed emitted mainly from this density region in Figure~\ref{fig:GW_SphHarm_both}.
On the other hand, we can see that in the region of $\rho > 10^{13}$~\gcm{}, where the kHz signal emanated, the $m = 1$ deformation develops only in the models with $1.5 \le \Omega_0 \le 3.5$~\rads{}. This supports our hypothesis that the kHz signal is associated with the low-\tw{} instability.
As a consequence, the strength of the kHz signal is correlated with the presence and amplitude of the $m = 1$ deformation in the high-density region of $\rho > 10^{13}$~\gcm{}.

Figure~\ref{fig:slice_Ynua} shows the mean energy of trapped electron anti-neutrinos for models with different initial angular velocities.
We can see that the mean energy of electron anti-neutrinos also exhibits prominent asymmetric distributions at $R_{13} \sim 20$~km in the models with $1.5 \le \Omega_0 \le 3.5$~\rads{}, which is consistent with the models for which the $m = 1$ density deformation and the kHz signal are present. 
This alignment once again highlights the correlation between the $m = 1$ density deformation and the neutrino distribution.
For neutrino treatments that include the $\mathcal{O}(v/c)$ terms, the density asymmetries can lead to an asymmetric neutrino pressure, potentially contributing to the development of the $m = 1$ density deformation in the PNS inner core.
Comparing Figure~\ref{fig:slice_dens} and Figure~\ref{fig:slice_Ynua}, we find that the phase of $m = 1$ density deformation leads the phase of the mean energy variation around $R_{13}$ (the system rotates counter-clockwise). This suggests that the high-density region produces more energetic neutrinos, subsequently heating the surrounding matter and facilitating expansion, which eventually results in a low-density region later.

A similar effect but with a lower modulation frequency has been proposed by \citet{2018MNRAS.475L..91T}.
They suggest that the neutrino emissions have a time modulation similar to the GW frequency from the low-\tw{} instability ($\sim 100-300$~Hz) and could be detectable by the Hyper-Kamiokande and the IceCube detectors. 
It is also expected that the asymmetric neutrino distributions associated with the kHz GW signal should produce noticeable time modulations on the neutrino emissions. However, in our current implementation of the IDSA, the free-streaming neutrinos are averaged over angles and therefore cannot be directly evaluated without additional approximations.

\subsection{Detectability of GW Signals} \label{sec:gw_obser}

\begin{figure}
	\plotone{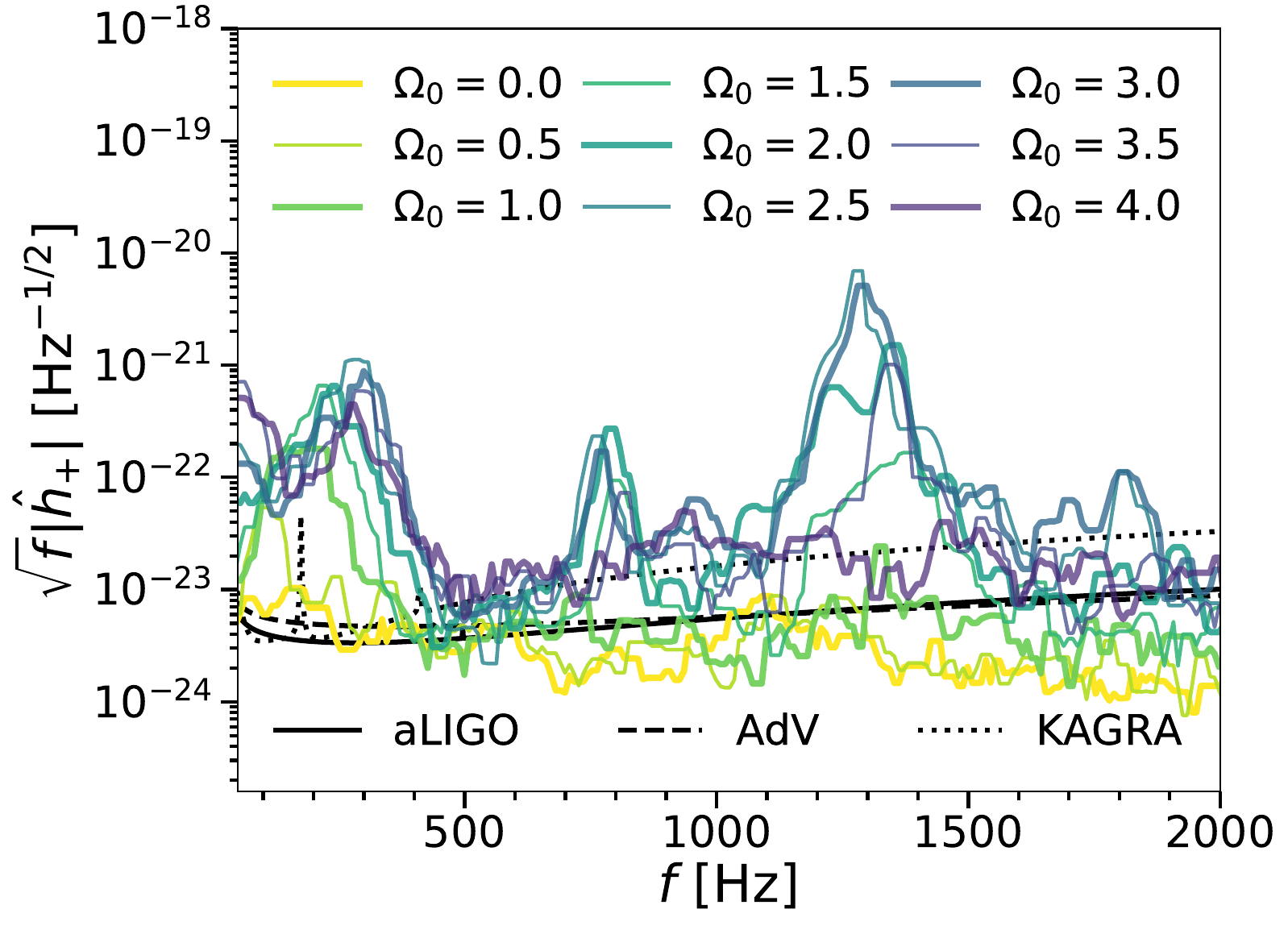}
	\caption{Amplitude spectral density of the plus mode of GW emissions between $t_\mathrm{pb} = -10$ and $100$~ms, for models using \sphynx{} with $675$k particles and various initial angular velocities, seen along the pole at a distance of $10$~kpc. The black solid, dashed, and dotted lines denote the sensitive curves of advanced LIGO (aLIGO), advanced Virgo (AdV), and KAGRA, respectively.}
	\label{fig:asd_omega}
\end{figure}

In this section, we discuss the detectability of the GW features from rapid-rotating CCSNe using the current ground-based GW detectors. 
Figure~\ref{fig:asd_omega} shows the amplitude spectral density (ASD) of the plus mode of GW emissions between $t_\mathrm{pb} = -10$ and $100$~ms in the \sphynx{} simulations with $675$k particles, seen along the pole and assumed at a distance $10$~kpc. 
Note that the bounce and ring-down signals are not present in this viewing angle which makes it easier to focus on the GW signals related to the low-\tw{} instability. 
The sensitivity curves of the GW detectors Advanced LIGO, Advanced Virgo, and KAGRA \citep{2020LRR....23....3A} are plotted as black lines for reference.
It is clear from Figure~\ref{fig:asd_omega} that both the $300$~Hz and kHz signals are detectable by the current ground-based GW detectors at our assumed source distance, and they are among the strongest signals in this time period. The peak frequencies of the $300$~Hz and kHz signals fall within narrow ranges between $210-300$~Hz and $1280-1350$~Hz, respectively, and are not sensitive to the initial angular velocity whenever these signals are present.

\begin{figure}
	\plotone{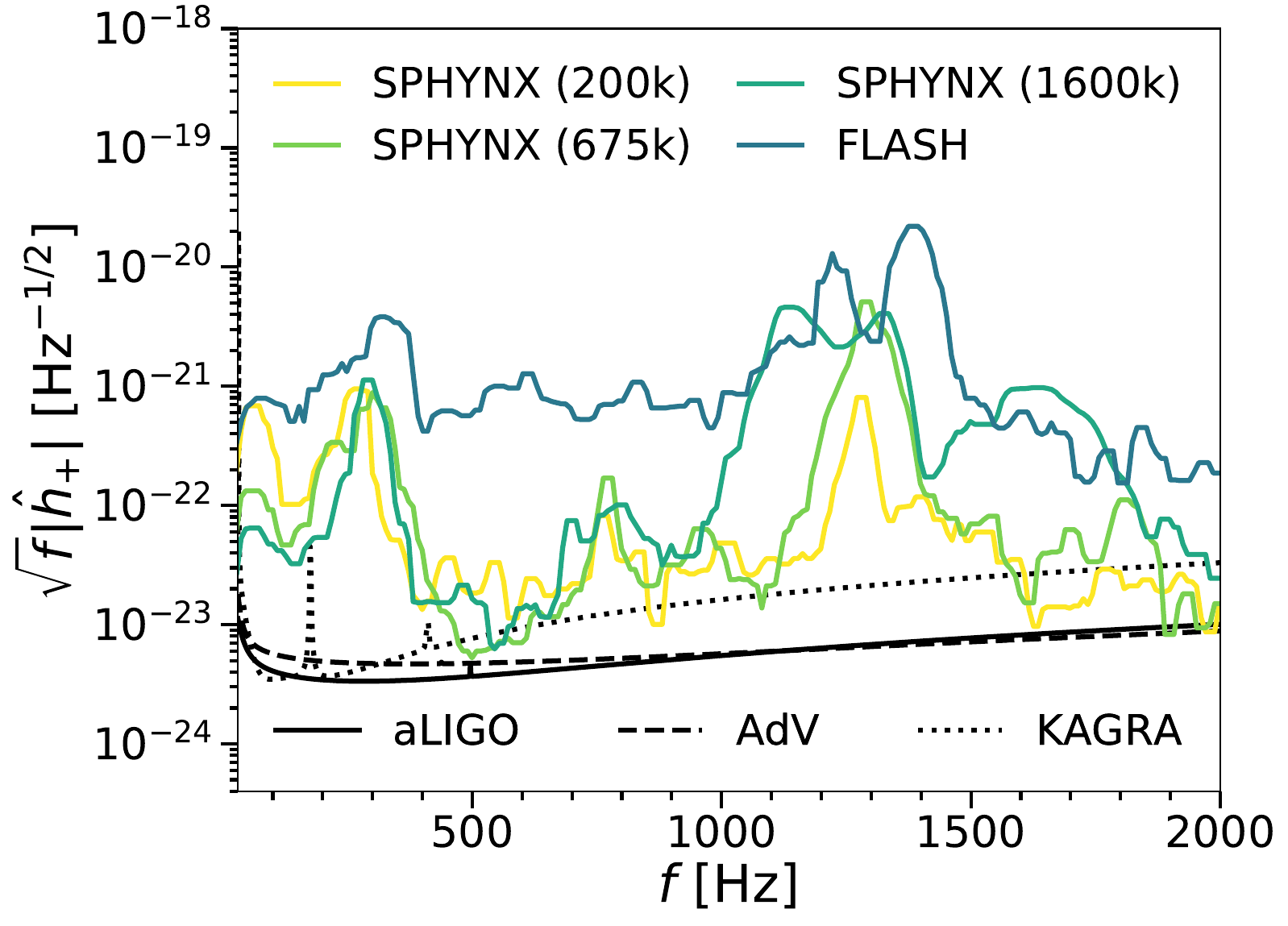}
	\caption{Similar to Figure~\ref{fig:asd_omega}, but for the $\Omega_0 = 3.0$ models using different codes and resolutions. For the comparison, only the GW emissions between $t_\mathrm{pb} = -10$ and $100$~ms are used for analysis.}
	\label{fig:asd_res}
\end{figure}

In addition, it is worth noting that there is one weaker but noticeable GW signal associated with the higher-order mode at around $800$~Hz, as discussed in Section~\ref{sec:gw_code}.
By examining Figures~\ref{fig:GW_sphynx_675k} and \ref{fig:asd_omega}, we can see that the peak frequency of this higher-order mode signal increases gradually after $t_\mathrm{pb} = 70-100$~ms, which leads to a secondary peak around $900-1000$~Hz in the ASD.
Among these GW features, the $300$~Hz signal is the most robust signal from the low-\tw{} instability and can be excited when the initial angular velocity is higher than $1.0$~\rads{}.
The kHz signal will appear when the initial angular velocity is within the range $1.5 \le \Omega_0 \le 3.5$~\rads{}. The cases with $\Omega_0 = 2.5$ and $3.0$~\rads{} have the strongest emissions, suggesting a resonance frequency at around $2.5-3.0$~\rads{}.
The higher-order mode signal, which peaked around $800$~Hz, appears in a similar range as the kHz signal ($1.5 \le \Omega_0 \le 3.5$~\rads{}) but does not show a clear resonance frequency.
Therefore, if we could detect the higher-order mode signal in addition to the $300$~Hz and kHz signals, these would provide additional constraints on narrowing down the initial angular velocity of a collapsar. 

In Figure~\ref{fig:asd_res}, we compare the ASD of GW emissions of \sphynx{} simulations with different numbers of particles, using $\Omega_0 = 3.0$~\rads{}.
We also plot the ASD of model F30 for comparison.
The ASD is evaluated within the time range between $t_\mathrm{pb} = -10$ and $100$~ms.
We can see that the GW emissions have a qualitatively similar spectral density distribution between the \sphynx{} simulations with $200$k (S30L), $675$k (S30), and $1600$k particles (S30H). 
However, model S30L shows a kHz signal that is one order of magnitude weaker in amplitude compared to models S30 and S30H, and the peak frequency is also $30 - 50$~Hz lower.
We consider that these differences are due to an underresolved PNS inner core in the lowest resolution model (S30L). 
Even with its low resolution, the S30L model displays an ASD similar to those of the S30 and S30H models, which have converged results. This is because the primary contributions to the GW emissions originate from the densest regions of the PNS, which have the highest spatial resolution in \sphynx{}. In addition, as long as the overall evolution is accurately described, the GW emissions can be adequately followed with integration over the entire domain, as discussed in Section~\ref{sec:method_gw}.

Comparing codes, the peak frequencies of the $300$~Hz and kHz low-\tw{} signals in the \flash{} simulation (F30) are qualitatively similar to those of the models S30 and S30H. Nevertheless, the GW emission in the F30 model is approximately one order of magnitude stronger in the signal amplitude than the results from \sphynx{}.
As pointed out in \cite{2021MNRAS.503.3552A}, the grid resolution in the post-shock region will affect the numerical damping of the convective cells and the activities of the g-mode around the PNS, which could explain the variation in the magnitudes of the GW emissions between \sphynx{} and \flash{}.

\section{SUMMARY \& CONCLUSIONS}
\label{sec:conclusion}

We have performed an analysis of the development of the low-\tw{} instability and associated GW emissions in the early postbounce phase of rotating CCSNe. To this end, we performed 3D hydrodynamical core-collapse simulations of a $20 M_\odot$ progenitor with different initial angular velocities ($\Omega_0$), which are parametrically added to the progenitor model. In this work, we computed models with $0 \le \Omega_0 \le 4$~\rads{} using the smoothed particle hydrodynamics code \sphynx{}, and models with $\Omega_0 = 2$ and $3$~\rads{} using the grid-based hydrodynamics code \flash{} for comparison.

Among our rotating models, the GW emissions exhibit two strong low-\tw{} signals after $20$~ms postbounce in both \sphynx{} and \flash{} simulations. Both signals are correlated with the $m = 1$ deformation induced by the low-\tw{} instability, with peak frequencies of about $300$~Hz and $1.3$~kHz (called kHz here). The $300$~Hz signal is present in models with $\Omega_0 \ge 1.0$~\rads{}, which mainly emanated from the region of $10^{11} \le \rho < 10^{13}$~\gcm{}. The peak frequency of the $300$~Hz signal is not sensitive to the initial angular velocity, the code used, or the spatial resolution. On the other hand, the kHz signal is present only in models with a narrower range of initial angular velocity, $1.5 \le \Omega_0 \le 3.5$~\rads{}, originates mainly in the region of $\rho \ge 10^{13}$~\gcm{}, and is highly associated with the asymmetric distribution of electron anti-neutrinos. The peak frequency of the kHz signal, once present, is not sensitive to the initial angular velocity, but is moderately affected by the dynamical evolution of the PNS inner core. 

In addition to the $300$~Hz and kHz signals, there is an additional weaker higher-order mode of GW emissions at around $800$~Hz, emanating mainly from the region of $10^{11} \le \rho < 10^{13}$~\gcm{}, in those of our models that developed the low-\tw{} instability. 
This higher-order signal is also correlated with the $m = 1$ deformation in the PNS, but its occurrence time is different from those of the $300$~Hz and kHz signals.
The range of initial angular velocity where the higher-order signal exists is similar to that of the kHz signal, around $1.5 \le \Omega_0 \le 3.5$~\rads{}. 
However, the peak frequency of this higher-order signal gradually increases to $900-1000$~Hz after $70-100$~ms postbounce, which leads to a secondary peak in the amplitude spectral density.
Therefore, the initial angular velocity of the CCSN progenitors can be inferred from the detection of the higher-order signal, and the $300$~Hz and kHz low-\tw{} signals.

We note that the GW features and their peak frequencies presented in this work can be dependent on the numerical methods and the physical models used, especially the kHz signal.
Furthermore, the kHz features associated with the low-\tw{} instability are typically originated from the asymmetric density distribution in the PNS inner core, where electron anti-neutrinos are largely produced and still coupled with the matter.
The density asymmetries in this region can induce an asymmetric neutrino distribution and consequently result in an asymmetric neutrino pressure. This, in turn, could facilitate the development of density deformation. A more in-depth stability analysis is necessary and will be one of our future works.

In this work, we have not considered the effects of the magnetic field, which can affect the dynamical evolution of the PNS and the explosion mechanism \citep[e.g.,][]{2020ApJ...896..102K, 2020MNRAS.499.4174M, 2020MNRAS.498L.109M, 2020MNRAS.492.4613O, 2021MNRAS.503.4942O, 2022MNRAS.509.3410R, 2023MNRAS.522.6070P}. In the presence of a strong magnetic field, the low-\tw{} instability could be suppressed \citep{2011MNRAS.413.2207F, 2014PhRvD..90j4014M}. The angular momentum transport driven by the magnetorotational instability can redistribute the rotational profile \citep{2018MNRAS.475..108B}, which in turn affects the development of the low-\tw{} instability and weakens the associated GW signals by an order of magnitude \citep{2023MNRAS.520.5622B}.
To obtain a more complete diagnostic of the angular momentum profile from the low-\tw{} signals, 3D magnetohydrodynamics simulations should be performed.

\acknowledgments{
We are grateful to the referee for a thoughtful report and useful suggestions that helped us improve the manuscript.
This work is supported by the National Center for Theoretical Sciences of Taiwan, the Ministry of Education (Higher Education Sprout Project NTU-112L104022), the National Science and Technology Council of Taiwan through grant NSTC 111-2112-M-007-037 and 112-2811-M-002-113, the Center for Informatics and Computation in Astronomy (CICA) at National Tsing Hua University through a grant from the Ministry of Education of Taiwan, and the Swiss Platform for Advanced Scientific Computing (PASC) projects SPH-EXA and SPH-EXA2: Optimizing Smooth Particle Hydrodynamics for Exascale Computing.
KCP is supported by the NSTC grant NSTC 111-2112-M-007-037 and 112-2112-M-007-040.
This work has also been carried out as part of the SKACH consortium through funding from SERI. Simulations and data analysis have been carried out on the {\tt Taiwania} supercomputer at the National Center for High-Performance Computing (NCHC) in Taiwan, on the CICA Cluster at the National Tsing Hua University, and on the scientific computing core facility \mbox{sciCORE} (http://scicore.unibas.ch/) at the University of Basel. 
Analysis and visualization of the simulation data were completed using the analysis toolkit yt \citep{2011ApJS..192....9T}.

\software{
SPHYNX \citep{2017A&A...606A..78C, 2022A&A...659A.175G},
FLASH \citep{2000ApJS..131..273F, 2008PhST..132a4046D}, 
Matplotlib \citep{2007CSE.....9...90H}, 
NumPy \citep{2011CSE....13b..22V}, 
PyCWT \citep{1998BAMS...79...61T}, 
SciPy \citep{2019zndo...3533894V}, 
yt \citep{2011ApJS..192....9T} 
}
}

\bibliographystyle{aasjournal}
\bibliography{reference}

\begin{thebibliography}{}
\expandafter\ifx\csname natexlab\endcsname\relax\def\natexlab#1{#1}\fi
\providecommand{\url}[1]{\href{#1}{#1}}
\providecommand{\dodoi}[1]{doi:~\href{https://doi.org/#1}{\nolinkurl{#1}}}
\providecommand{\doeprint}[1]{\href{https://ascl.net/#1}{\nolinkurl{https://ascl.net/#1}}}
\providecommand{\doarXiv}[1]{\href{https://arxiv.org/abs/#1}{\nolinkurl{https://arxiv.org/abs/#1}}}

\bibitem[{{Abbott} {et~al.}(2020){Abbott}, {Abbott}, {Abbott}, {Abraham},
  {Acernese}, {Ackley}, {Adams}, {Adya}, {Affeldt}, {Agathos}, \&
  et~al.}]{2020LRR....23....3A}
{Abbott}, B.~P., {Abbott}, R., {Abbott}, T.~D., {et~al.} 2020, Living Reviews
  in Relativity, 23, 3, \dodoi{10.1007/s41114-020-00026-9}

\bibitem[{{Abdikamalov} {et~al.}(2014){Abdikamalov}, {Gossan}, {DeMaio}, \&
  {Ott}}]{2014PhRvD..90d4001A}
{Abdikamalov}, E., {Gossan}, S., {DeMaio}, A.~M., \& {Ott}, C.~D. 2014, \prd,
  90, 044001, \dodoi{10.1103/PhysRevD.90.044001}

\bibitem[{{Abdikamalov} {et~al.}(2022){Abdikamalov}, {Pagliaroli}, \&
  {Radice}}]{2022hgwa.bookE..21A}
{Abdikamalov}, E., {Pagliaroli}, G., \& {Radice}, D. 2022, in Handbook of
  Gravitational Wave Astronomy. Edited by C. Bambi, 21,
  \dodoi{10.1007/978-981-15-4702-7_21-1}

\bibitem[{{Andresen} {et~al.}(2021){Andresen}, {Glas}, \&
  {Janka}}]{2021MNRAS.503.3552A}
{Andresen}, H., {Glas}, R., \& {Janka}, H.~T. 2021, \mnras, 503, 3552,
  \dodoi{10.1093/mnras/stab675}

\bibitem[{{Andresen} {et~al.}(2019){Andresen}, {M{\"u}ller}, {Janka}, {Summa},
  {Gill}, \& {Zanolin}}]{2019MNRAS.486.2238A}
{Andresen}, H., {M{\"u}ller}, E., {Janka}, H.~T., {et~al.} 2019, \mnras, 486,
  2238, \dodoi{10.1093/mnras/stz990}

\bibitem[{{Blondin} {et~al.}(2003){Blondin}, {Mezzacappa}, \&
  {DeMarino}}]{2003ApJ...584..971B}
{Blondin}, J.~M., {Mezzacappa}, A., \& {DeMarino}, C. 2003, \apj, 584, 971,
  \dodoi{10.1086/345812}

\bibitem[{{Bugli} {et~al.}(2023){Bugli}, {Guilet}, {Foglizzo}, \&
  {Obergaulinger}}]{2023MNRAS.520.5622B}
{Bugli}, M., {Guilet}, J., {Foglizzo}, T., \& {Obergaulinger}, M. 2023, \mnras,
  520, 5622, \dodoi{10.1093/mnras/stad496}

\bibitem[{{Bugli} {et~al.}(2018){Bugli}, {Guilet}, {M{\"u}ller}, {Del Zanna},
  {Bucciantini}, \& {Montero}}]{2018MNRAS.475..108B}
{Bugli}, M., {Guilet}, J., {M{\"u}ller}, E., {et~al.} 2018, \mnras, 475, 108,
  \dodoi{10.1093/mnras/stx3158}

\bibitem[{{Buras} {et~al.}(2006){Buras}, {Rampp}, {Janka}, \&
  {Kifonidis}}]{2006A&A...447.1049B}
{Buras}, R., {Rampp}, M., {Janka}, H.~T., \& {Kifonidis}, K. 2006, \aap, 447,
  1049, \dodoi{10.1051/0004-6361:20053783}

\bibitem[{{Burrows} {et~al.}(2012){Burrows}, {Dolence}, \&
  {Murphy}}]{2012ApJ...759....5B}
{Burrows}, A., {Dolence}, J.~C., \& {Murphy}, J.~W. 2012, \apj, 759, 5,
  \dodoi{10.1088/0004-637X/759/1/5}

\bibitem[{{Burrows} {et~al.}(2019){Burrows}, {Radice}, \&
  {Vartanyan}}]{2019MNRAS.485.3153B}
{Burrows}, A., {Radice}, D., \& {Vartanyan}, D. 2019, \mnras, 485, 3153,
  \dodoi{10.1093/mnras/stz543}

\bibitem[{{Burrows} \& {Vartanyan}(2021)}]{2021Natur.589...29B}
{Burrows}, A., \& {Vartanyan}, D. 2021, \nat, 589, 29,
  \dodoi{10.1038/s41586-020-03059-w}

\bibitem[{{Cabez{\'o}n} {et~al.}(2017){Cabez{\'o}n}, {Garc{\'\i}a-Senz}, \&
  {Figueira}}]{2017A&A...606A..78C}
{Cabez{\'o}n}, R.~M., {Garc{\'\i}a-Senz}, D., \& {Figueira}, J. 2017, \aap,
  606, A78, \dodoi{10.1051/0004-6361/201630208}

\bibitem[{{Cabez{\'o}n} {et~al.}(2008){Cabez{\'o}n}, {Garc{\'\i}a-Senz}, \&
  {Rela{\~n}o}}]{2008JCoPh.227.8523C}
{Cabez{\'o}n}, R.~M., {Garc{\'\i}a-Senz}, D., \& {Rela{\~n}o}, A. 2008, Journal
  of Computational Physics, 227, 8523, \dodoi{10.1016/j.jcp.2008.06.014}

\bibitem[{{Cabez{\'o}n} {et~al.}(2018){Cabez{\'o}n}, {Pan}, {Liebend{\"o}rfer},
  {Kuroda}, {Ebinger}, {Heinimann}, {Perego}, \&
  {Thielemann}}]{2018A&A...619A.118C}
{Cabez{\'o}n}, R.~M., {Pan}, K.-C., {Liebend{\"o}rfer}, M., {et~al.} 2018,
  \aap, 619, A118, \dodoi{10.1051/0004-6361/201833705}

\bibitem[{{Centrella} \& {McMillan}(1993)}]{1993ApJ...416..719C}
{Centrella}, J.~M., \& {McMillan}, S. L.~W. 1993, \apj, 416, 719,
  \dodoi{10.1086/173272}

\bibitem[{{Centrella} {et~al.}(2001){Centrella}, {New}, {Lowe}, \&
  {Brown}}]{2001ApJ...550L.193C}
{Centrella}, J.~M., {New}, K. C.~B., {Lowe}, L.~L., \& {Brown}, J.~D. 2001,
  \apjl, 550, L193, \dodoi{10.1086/319634}

\bibitem[{{Dubey} {et~al.}(2008){Dubey}, {Reid}, \&
  {Fisher}}]{2008PhST..132a4046D}
{Dubey}, A., {Reid}, L.~B., \& {Fisher}, R. 2008, Physica Scripta Volume T,
  132, 014046, \dodoi{10.1088/0031-8949/2008/T132/014046}

\bibitem[{{Finn} \& {Evans}(1990)}]{1990ApJ...351..588F}
{Finn}, L.~S., \& {Evans}, C.~R. 1990, \apj, 351, 588, \dodoi{10.1086/168497}

\bibitem[{{Fryxell} {et~al.}(2000){Fryxell}, {Olson}, {Ricker}, {Timmes},
  {Zingale}, {Lamb}, {MacNeice}, {Rosner}, {Truran}, \&
  {Tufo}}]{2000ApJS..131..273F}
{Fryxell}, B., {Olson}, K., {Ricker}, P., {et~al.} 2000, \apjs, 131, 273,
  \dodoi{10.1086/317361}

\bibitem[{{Fu} \& {Lai}(2011)}]{2011MNRAS.413.2207F}
{Fu}, W., \& {Lai}, D. 2011, \mnras, 413, 2207,
  \dodoi{10.1111/j.1365-2966.2011.18296.x}

\bibitem[{{Garc{\'\i}a-Senz} {et~al.}(2012){Garc{\'\i}a-Senz}, {Cabez{\'o}n},
  \& {Escart{\'\i}n}}]{2012A&A...538A...9G}
{Garc{\'\i}a-Senz}, D., {Cabez{\'o}n}, R.~M., \& {Escart{\'\i}n}, J.~A. 2012,
  \aap, 538, A9, \dodoi{10.1051/0004-6361/201117939}

\bibitem[{{Garc{\'\i}a-Senz} {et~al.}(2022){Garc{\'\i}a-Senz}, {Cabez{\'o}n},
  \& {Escart{\'\i}n}}]{2022A&A...659A.175G}
---. 2022, \aap, 659, A175, \dodoi{10.1051/0004-6361/202141877}

\bibitem[{{Hopkins}(2013)}]{2013MNRAS.428.2840H}
{Hopkins}, P.~F. 2013, \mnras, 428, 2840, \dodoi{10.1093/mnras/sts210}

\bibitem[{{Hunter}(2007)}]{2007CSE.....9...90H}
{Hunter}, J.~D. 2007, Computing in Science and Engineering, 9, 90,
  \dodoi{10.1109/MCSE.2007.55}

\bibitem[{{Ivanov} \& {Fern{\'a}ndez}(2021)}]{2021ApJ...911....6I}
{Ivanov}, M., \& {Fern{\'a}ndez}, R. 2021, \apj, 911, 6,
  \dodoi{10.3847/1538-4357/abe59e}

\bibitem[{{Jardine} {et~al.}(2022){Jardine}, {Powell}, \&
  {M{\"u}ller}}]{2022MNRAS.510.5535J}
{Jardine}, R., {Powell}, J., \& {M{\"u}ller}, B. 2022, \mnras, 510, 5535,
  \dodoi{10.1093/mnras/stab3763}

\bibitem[{{Kuroda} {et~al.}(2020){Kuroda}, {Arcones}, {Takiwaki}, \&
  {Kotake}}]{2020ApJ...896..102K}
{Kuroda}, T., {Arcones}, A., {Takiwaki}, T., \& {Kotake}, K. 2020, \apj, 896,
  102, \dodoi{10.3847/1538-4357/ab9308}

\bibitem[{{Kuroda} {et~al.}(2016){Kuroda}, {Kotake}, \&
  {Takiwaki}}]{2016ApJ...829L..14K}
{Kuroda}, T., {Kotake}, K., \& {Takiwaki}, T. 2016, \apjl, 829, L14,
  \dodoi{10.3847/2041-8205/829/1/L14}

\bibitem[{{Kuroda} {et~al.}(2014){Kuroda}, {Takiwaki}, \&
  {Kotake}}]{2014PhRvD..89d4011K}
{Kuroda}, T., {Takiwaki}, T., \& {Kotake}, K. 2014, \prd, 89, 044011,
  \dodoi{10.1103/PhysRevD.89.044011}

\bibitem[{{Lattimer} \& {Swesty}(1991)}]{1991NuPhA.535..331L}
{Lattimer}, J.~M., \& {Swesty}, D.~F. 1991, \nphysa, 535, 331,
  \dodoi{10.1016/0375-9474(91)90452-C}

\bibitem[{{Ledoux}(1947)}]{1947ApJ...105..305L}
{Ledoux}, P. 1947, \apj, 105, 305, \dodoi{10.1086/144905}

\bibitem[{{Liebend{\"o}rfer}(2005)}]{2005ApJ...633.1042L}
{Liebend{\"o}rfer}, M. 2005, \apj, 633, 1042, \dodoi{10.1086/466517}

\bibitem[{{Liebend{\"o}rfer} {et~al.}(2009){Liebend{\"o}rfer}, {Whitehouse}, \&
  {Fischer}}]{2009ApJ...698.1174L}
{Liebend{\"o}rfer}, M., {Whitehouse}, S.~C., \& {Fischer}, T. 2009, \apj, 698,
  1174, \dodoi{10.1088/0004-637X/698/2/1174}

\bibitem[{{Liu} {et~al.}(2007){Liu}, {San Liang}, \&
  {Weisberg}}]{2007JAtOT..24.2093L}
{Liu}, Y., {San Liang}, X., \& {Weisberg}, R.~H. 2007, Journal of Atmospheric
  and Oceanic Technology, 24, 2093, \dodoi{10.1175/2007JTECHO511.1}

\bibitem[{{Marek} {et~al.}(2006){Marek}, {Dimmelmeier}, {Janka}, {M{\"u}ller},
  \& {Buras}}]{2006A&A...445..273M}
{Marek}, A., {Dimmelmeier}, H., {Janka}, H.~T., {M{\"u}ller}, E., \& {Buras},
  R. 2006, \aap, 445, 273, \dodoi{10.1051/0004-6361:20052840}

\bibitem[{{Matsumoto} {et~al.}(2020){Matsumoto}, {Takiwaki}, {Kotake},
  {Asahina}, \& {Takahashi}}]{2020MNRAS.499.4174M}
{Matsumoto}, J., {Takiwaki}, T., {Kotake}, K., {Asahina}, Y., \& {Takahashi},
  H.~R. 2020, \mnras, 499, 4174, \dodoi{10.1093/mnras/staa3095}

\bibitem[{{Melson} {et~al.}(2015){Melson}, {Janka}, \&
  {Marek}}]{2015ApJ...801L..24M}
{Melson}, T., {Janka}, H.-T., \& {Marek}, A. 2015, \apjl, 801, L24,
  \dodoi{10.1088/2041-8205/801/2/L24}

\bibitem[{{Mezzacappa} {et~al.}(2020){Mezzacappa}, {Marronetti}, {Landfield},
  {Lentz}, {Yakunin}, {Bruenn}, {Hix}, {Messer}, {Endeve}, {Blondin}, \&
  {Harris}}]{2020PhRvD.102b3027M}
{Mezzacappa}, A., {Marronetti}, P., {Landfield}, R.~E., {et~al.} 2020, \prd,
  102, 023027, \dodoi{10.1103/PhysRevD.102.023027}

\bibitem[{{Mezzacappa} {et~al.}(2023){Mezzacappa}, {Marronetti}, {Landfield},
  {Lentz}, {Murphy}, {Hix}, {Harris}, {Bruenn}, {Blondin}, {Bronson Messer},
  {Casanova}, \& {Kronzer}}]{2023PhRvD.107d3008M}
---. 2023, \prd, 107, 043008, \dodoi{10.1103/PhysRevD.107.043008}

\bibitem[{{Monaghan}(2005)}]{2005RPPh...68.1703M}
{Monaghan}, J.~J. 2005, Reports on Progress in Physics, 68, 1703,
  \dodoi{10.1088/0034-4885/68/8/R01}

\bibitem[{{Morozova} {et~al.}(2018){Morozova}, {Radice}, {Burrows}, \&
  {Vartanyan}}]{2018ApJ...861...10M}
{Morozova}, V., {Radice}, D., {Burrows}, A., \& {Vartanyan}, D. 2018, \apj,
  861, 10, \dodoi{10.3847/1538-4357/aac5f1}

\bibitem[{{Muhlberger} {et~al.}(2014){Muhlberger}, {Nouri}, {Duez}, {Foucart},
  {Kidder}, {Ott}, {Scheel}, {Szil{\'a}gyi}, \&
  {Teukolsky}}]{2014PhRvD..90j4014M}
{Muhlberger}, C.~D., {Nouri}, F.~H., {Duez}, M.~D., {et~al.} 2014, \prd, 90,
  104014, \dodoi{10.1103/PhysRevD.90.104014}

\bibitem[{{M{\"u}ller} \& {Varma}(2020)}]{2020MNRAS.498L.109M}
{M{\"u}ller}, B., \& {Varma}, V. 2020, \mnras, 498, L109,
  \dodoi{10.1093/mnrasl/slaa137}

\bibitem[{{Obergaulinger} \& {Aloy}(2020)}]{2020MNRAS.492.4613O}
{Obergaulinger}, M., \& {Aloy}, M.~{\'A}. 2020, \mnras, 492, 4613,
  \dodoi{10.1093/mnras/staa096}

\bibitem[{{Obergaulinger} \& {Aloy}(2021)}]{2021MNRAS.503.4942O}
---. 2021, \mnras, 503, 4942, \dodoi{10.1093/mnras/stab295}

\bibitem[{{O'Connor} \& {Ott}(2010)}]{2010CQGra..27k4103O}
{O'Connor}, E., \& {Ott}, C.~D. 2010, Classical and Quantum Gravity, 27,
  114103, \dodoi{10.1088/0264-9381/27/11/114103}

\bibitem[{{O'Connor} \& {Couch}(2018)}]{2018ApJ...865...81O}
{O'Connor}, E.~P., \& {Couch}, S.~M. 2018, \apj, 865, 81,
  \dodoi{10.3847/1538-4357/aadcf7}

\bibitem[{{Oohara} {et~al.}(1997){Oohara}, {Nakamura}, \&
  {Shibata}}]{1997PThPS.128..183O}
{Oohara}, K.-i., {Nakamura}, T., \& {Shibata}, M. 1997, Progress of Theoretical
  Physics Supplement, 128, 183, \dodoi{10.1143/PTPS.128.183}

\bibitem[{{Ott} {et~al.}(2007){Ott}, {Dimmelmeier}, {Marek}, {Janka}, {Hawke},
  {Zink}, \& {Schnetter}}]{2007PhRvL..98z1101O}
{Ott}, C.~D., {Dimmelmeier}, H., {Marek}, A., {et~al.} 2007, \prl, 98, 261101,
  \dodoi{10.1103/PhysRevLett.98.261101}

\bibitem[{{Ott} {et~al.}(2005){Ott}, {Ou}, {Tohline}, \&
  {Burrows}}]{2005ApJ...625L.119O}
{Ott}, C.~D., {Ou}, S., {Tohline}, J.~E., \& {Burrows}, A. 2005, \apjl, 625,
  L119, \dodoi{10.1086/431305}

\bibitem[{{Ott} {et~al.}(2013){Ott}, {Abdikamalov}, {M{\"o}sta}, {Haas},
  {Drasco}, {O'Connor}, {Reisswig}, {Meakin}, \&
  {Schnetter}}]{2013ApJ...768..115O}
{Ott}, C.~D., {Abdikamalov}, E., {M{\"o}sta}, P., {et~al.} 2013, \apj, 768,
  115, \dodoi{10.1088/0004-637X/768/2/115}

\bibitem[{{Pajkos} {et~al.}(2019){Pajkos}, {Couch}, {Pan}, \&
  {O'Connor}}]{2019ApJ...878...13P}
{Pajkos}, M.~A., {Couch}, S.~M., {Pan}, K.-C., \& {O'Connor}, E.~P. 2019, \apj,
  878, 13, \dodoi{10.3847/1538-4357/ab1de2}

\bibitem[{{Pajkos} {et~al.}(2021){Pajkos}, {Warren}, {Couch}, {O'Connor}, \&
  {Pan}}]{2021ApJ...914...80P}
{Pajkos}, M.~A., {Warren}, M.~L., {Couch}, S.~M., {O'Connor}, E.~P., \& {Pan},
  K.-C. 2021, \apj, 914, 80, \dodoi{10.3847/1538-4357/abfb65}

\bibitem[{{Pan} {et~al.}(2018){Pan}, {Liebend{\"o}rfer}, {Couch}, \&
  {Thielemann}}]{2018ApJ...857...13P}
{Pan}, K.-C., {Liebend{\"o}rfer}, M., {Couch}, S.~M., \& {Thielemann}, F.-K.
  2018, \apj, 857, 13, \dodoi{10.3847/1538-4357/aab71d}

\bibitem[{{Pan} {et~al.}(2021){Pan}, {Liebend{\"o}rfer}, {Couch}, \&
  {Thielemann}}]{2021ApJ...914..140P}
---. 2021, \apj, 914, 140, \dodoi{10.3847/1538-4357/abfb05}

\bibitem[{{Pan} {et~al.}(2016){Pan}, {Liebend{\"o}rfer}, {Hempel}, \&
  {Thielemann}}]{2016ApJ...817...72P}
{Pan}, K.-C., {Liebend{\"o}rfer}, M., {Hempel}, M., \& {Thielemann}, F.-K.
  2016, \apj, 817, 72, \dodoi{10.3847/0004-637X/817/1/72}

\bibitem[{{Pan} {et~al.}(2019){Pan}, {Mattes}, {O'Connor}, {Couch}, {Perego},
  \& {Arcones}}]{2019JPhG...46a4001P}
{Pan}, K.-C., {Mattes}, C., {O'Connor}, E.~P., {et~al.} 2019, Journal of
  Physics G Nuclear Physics, 46, 014001, \dodoi{10.1088/1361-6471/aaed51}

\bibitem[{{Perego} {et~al.}(2014){Perego}, {Gafton}, {Cabez{\'o}n}, {Rosswog},
  \& {Liebend{\"o}rfer}}]{2014A&A...568A..11P}
{Perego}, A., {Gafton}, E., {Cabez{\'o}n}, R., {Rosswog}, S., \&
  {Liebend{\"o}rfer}, M. 2014, \aap, 568, A11,
  \dodoi{10.1051/0004-6361/201423755}

\bibitem[{{Powell} \& {M{\"u}ller}(2020)}]{2020MNRAS.494.4665P}
{Powell}, J., \& {M{\"u}ller}, B. 2020, \mnras, 494, 4665,
  \dodoi{10.1093/mnras/staa1048}

\bibitem[{{Powell} {et~al.}(2023){Powell}, {M{\"u}ller}, {Aguilera-Dena}, \&
  {Langer}}]{2023MNRAS.522.6070P}
{Powell}, J., {M{\"u}ller}, B., {Aguilera-Dena}, D.~R., \& {Langer}, N. 2023,
  \mnras, 522, 6070, \dodoi{10.1093/mnras/stad1292}

\bibitem[{{Powell} {et~al.}(2021){Powell}, {M{\"u}ller}, \&
  {Heger}}]{2021MNRAS.503.2108P}
{Powell}, J., {M{\"u}ller}, B., \& {Heger}, A. 2021, \mnras, 503, 2108,
  \dodoi{10.1093/mnras/stab614}

\bibitem[{{Rahman} {et~al.}(2022){Rahman}, {Janka}, {Stockinger}, \&
  {Woosley}}]{2022MNRAS.512.4503R}
{Rahman}, N., {Janka}, H.~T., {Stockinger}, G., \& {Woosley}, S.~E. 2022,
  \mnras, 512, 4503, \dodoi{10.1093/mnras/stac758}

\bibitem[{{Raynaud} {et~al.}(2022){Raynaud}, {Cerd{\'a}-Dur{\'a}n}, \&
  {Guilet}}]{2022MNRAS.509.3410R}
{Raynaud}, R., {Cerd{\'a}-Dur{\'a}n}, P., \& {Guilet}, J. 2022, \mnras, 509,
  3410, \dodoi{10.1093/mnras/stab3109}

\bibitem[{{Read} {et~al.}(2010){Read}, {Hayfield}, \&
  {Agertz}}]{2010MNRAS.405.1513R}
{Read}, J.~I., {Hayfield}, T., \& {Agertz}, O. 2010, \mnras, 405, 1513,
  \dodoi{10.1111/j.1365-2966.2010.16577.x}

\bibitem[{{Richers} {et~al.}(2017){Richers}, {Ott}, {Abdikamalov}, {O'Connor},
  \& {Sullivan}}]{2017PhRvD..95f3019R}
{Richers}, S., {Ott}, C.~D., {Abdikamalov}, E., {O'Connor}, E., \& {Sullivan},
  C. 2017, \prd, 95, 063019, \dodoi{10.1103/PhysRevD.95.063019}

\bibitem[{{Rosswog}(2015)}]{2015MNRAS.448.3628R}
{Rosswog}, S. 2015, \mnras, 448, 3628, \dodoi{10.1093/mnras/stv225}

\bibitem[{{Saijo} {et~al.}(2003){Saijo}, {Baumgarte}, \&
  {Shapiro}}]{2003ApJ...595..352S}
{Saijo}, M., {Baumgarte}, T.~W., \& {Shapiro}, S.~L. 2003, \apj, 595, 352,
  \dodoi{10.1086/377334}

\bibitem[{{Saijo} \& {Yoshida}(2006)}]{2006MNRAS.368.1429S}
{Saijo}, M., \& {Yoshida}, S. 2006, \mnras, 368, 1429,
  \dodoi{10.1111/j.1365-2966.2006.10229.x}

\bibitem[{{Saitoh} \& {Makino}(2013)}]{2013ApJ...768...44S}
{Saitoh}, T.~R., \& {Makino}, J. 2013, \apj, 768, 44,
  \dodoi{10.1088/0004-637X/768/1/44}

\bibitem[{{Scheidegger} {et~al.}(2008){Scheidegger}, {Fischer}, {Whitehouse},
  \& {Liebend{\"o}rfer}}]{2008A&A...490..231S}
{Scheidegger}, S., {Fischer}, T., {Whitehouse}, S.~C., \& {Liebend{\"o}rfer},
  M. 2008, \aap, 490, 231, \dodoi{10.1051/0004-6361:20078577}

\bibitem[{{Scheidegger} {et~al.}(2010){Scheidegger}, {K{\"a}ppeli},
  {Whitehouse}, {Fischer}, \& {Liebend{\"o}rfer}}]{2010A&A...514A..51S}
{Scheidegger}, S., {K{\"a}ppeli}, R., {Whitehouse}, S.~C., {Fischer}, T., \&
  {Liebend{\"o}rfer}, M. 2010, \aap, 514, A51,
  \dodoi{10.1051/0004-6361/200913220}

\bibitem[{{Shibagaki} {et~al.}(2020){Shibagaki}, {Kuroda}, {Kotake}, \&
  {Takiwaki}}]{2020MNRAS.493L.138S}
{Shibagaki}, S., {Kuroda}, T., {Kotake}, K., \& {Takiwaki}, T. 2020, \mnras,
  493, L138, \dodoi{10.1093/mnrasl/slaa021}

\bibitem[{{Shibagaki} {et~al.}(2021){Shibagaki}, {Kuroda}, {Kotake}, \&
  {Takiwaki}}]{2021MNRAS.502.3066S}
---. 2021, \mnras, 502, 3066, \dodoi{10.1093/mnras/stab228}

\bibitem[{{Sotani} {et~al.}(2021){Sotani}, {Takiwaki}, \&
  {Togashi}}]{2021PhRvD.104l3009S}
{Sotani}, H., {Takiwaki}, T., \& {Togashi}, H. 2021, \prd, 104, 123009,
  \dodoi{10.1103/PhysRevD.104.123009}

\bibitem[{{Szczepa{\'n}czyk} {et~al.}(2021){Szczepa{\'n}czyk}, {Antelis},
  {Benjamin}, {Cavagli{\`a}}, {Gondek-Rosi{\'n}ska}, {Hansen}, {Klimenko},
  {Morales}, {Moreno}, {Mukherjee}, {Nurbek}, {Powell}, {Singh},
  {Sitmukhambetov}, {Szewczyk}, {Valdez}, {Vedovato}, {Westhouse}, {Zanolin},
  \& {Zheng}}]{2021PhRvD.104j2002S}
{Szczepa{\'n}czyk}, M.~J., {Antelis}, J.~M., {Benjamin}, M., {et~al.} 2021,
  \prd, 104, 102002, \dodoi{10.1103/PhysRevD.104.102002}

\bibitem[{{Takiwaki} \& {Kotake}(2018)}]{2018MNRAS.475L..91T}
{Takiwaki}, T., \& {Kotake}, K. 2018, \mnras, 475, L91,
  \dodoi{10.1093/mnrasl/sly008}

\bibitem[{{Takiwaki} {et~al.}(2021){Takiwaki}, {Kotake}, \&
  {Foglizzo}}]{2021MNRAS.508..966T}
{Takiwaki}, T., {Kotake}, K., \& {Foglizzo}, T. 2021, \mnras, 508, 966,
  \dodoi{10.1093/mnras/stab2607}

\bibitem[{{Takiwaki} {et~al.}(2016){Takiwaki}, {Kotake}, \&
  {Suwa}}]{2016MNRAS.461L.112T}
{Takiwaki}, T., {Kotake}, K., \& {Suwa}, Y. 2016, \mnras, 461, L112,
  \dodoi{10.1093/mnrasl/slw105}

\bibitem[{{Torrence} \& {Compo}(1998)}]{1998BAMS...79...61T}
{Torrence}, C., \& {Compo}, G.~P. 1998, Bulletin of the American Meteorological
  Society, 79, 61, \dodoi{10.1175/1520-0477(1998)079<0061:APGTWA>2.0.CO;2}

\bibitem[{{Torres-Forn{\'e}} {et~al.}(2018){Torres-Forn{\'e}},
  {Cerd{\'a}-Dur{\'a}n}, {Passamonti}, \& {Font}}]{2018MNRAS.474.5272T}
{Torres-Forn{\'e}}, A., {Cerd{\'a}-Dur{\'a}n}, P., {Passamonti}, A., \& {Font},
  J.~A. 2018, \mnras, 474, 5272, \dodoi{10.1093/mnras/stx3067}

\bibitem[{{Turk} {et~al.}(2011){Turk}, {Smith}, {Oishi}, {Skory}, {Skillman},
  {Abel}, \& {Norman}}]{2011ApJS..192....9T}
{Turk}, M.~J., {Smith}, B.~D., {Oishi}, J.~S., {et~al.} 2011, \apjs, 192, 9,
  \dodoi{10.1088/0067-0049/192/1/9}

\bibitem[{{van der Walt} {et~al.}(2011){van der Walt}, {Colbert}, \&
  {Varoquaux}}]{2011CSE....13b..22V}
{van der Walt}, S., {Colbert}, S.~C., \& {Varoquaux}, G. 2011, Computing in
  Science and Engineering, 13, 22, \dodoi{10.1109/MCSE.2011.37}

\bibitem[{{Vartanyan} \& {Burrows}(2020)}]{2020ApJ...901..108V}
{Vartanyan}, D., \& {Burrows}, A. 2020, \apj, 901, 108,
  \dodoi{10.3847/1538-4357/abafac}

\bibitem[{{Virtanen} {et~al.}(2019){Virtanen}, {Gommers}, {Burovski},
  {Oliphant}, {Cournapeau}, {Weckesser}, {alexbrc}, {Peterson}, {Wilson},
  {endolith}, {Mayorov}, {van der Walt}, {Laxalde}, {Haberland}, {Brett},
  {Nelson}, {Millman}, {Reddy}, {Larson}, {Lars}, {Polat}, {eric-jones},
  {Carey}, {Kern}, {Moore}, {Leslie}, {Perktold}, {Feng}, {Bharti}, \&
  {Vanderplas}}]{2019zndo...3533894V}
{Virtanen}, P., {Gommers}, R., {Burovski}, E., {et~al.} 2019, {scipy/scipy:
  SciPy 1.3.2}, v1.3.2, Zenodo,  Zenodo, \dodoi{10.5281/zenodo.3533894}

\bibitem[{{Watts} {et~al.}(2005){Watts}, {Andersson}, \&
  {Jones}}]{2005ApJ...618L..37W}
{Watts}, A.~L., {Andersson}, N., \& {Jones}, D.~I. 2005, \apjl, 618, L37,
  \dodoi{10.1086/427653}

\bibitem[{{Woosley} \& {Heger}(2007)}]{2007PhR...442..269W}
{Woosley}, S.~E., \& {Heger}, A. 2007, \physrep, 442, 269,
  \dodoi{10.1016/j.physrep.2007.02.009}

\end{thebibliography}

\end{CJK*}
\end{document}